\documentclass[10pt]{article}
\usepackage[top=1.8cm, bottom=1.8cm, left=1.8cm, right=1.8cm]{geometry}
\usepackage{helvet}  %
\usepackage{courier}  %
\usepackage[hyphens]{url}  %
\usepackage[keeplastbox]{flushend}  %
\usepackage{graphicx}  %
\newif\ifcomment
 \commentfalse

\usepackage{xcolor}
\usepackage{booktabs}
\usepackage{graphicx}
\usepackage{paralist}
\usepackage[small,bf]{caption}
\usepackage{subfigure}
\usepackage{times}
\usepackage[hang,flushmargin]{footmisc}

\usepackage[compact]{titlesec}
\titlespacing*{\section}{0pt}{*4}{4pt} 
\titlespacing{\subsection}{0pt}{*3}{3pt}

\usepackage[bookmarks=true,%
            bookmarksnumbered=true,%
            colorlinks=true,%
            linkcolor=linkcol,%
            citecolor=citecol,%
            urlcolor=urlcol,%
            hypertexnames=true,%
            pdfpagelabels]%
      {hyperref}

\definecolor{linkcol}{rgb}{0,0,0.5}
\definecolor{citecol}{rgb}{0,0.5,0.3}
\definecolor{urlcol}{rgb}{0.3,0,0}

\usepackage{xspace}

\usepackage{todonotes} %

\usepackage{ifthen}
\newif\ifshort
\shortfalse %
\ifshort
  \newcommand{\isShort}{true}
\else
  \newcommand{\isShort}{false}
\fi
\newcommand{\shortVer}[1]{\ifthenelse{\equal{\isShort}{true}}{{#1}}{}}
\newcommand{\longVer}[1]{\ifthenelse{\equal{\isShort}{false}}{{#1}}{}}

\ifcomment
\newcommand{\XXX}[2]{{\bf \textcolor{blue}{#1: #2}}}
\newcommand{\XXXR}[2]{{\bf \textcolor{red}{#1: #2}}}
\newcommand{\jbnote}[1]{{\bf \textcolor{magenta}{JB: #1}}}
\newcommand{\msnote}[1]{{\bf \textcolor{magenta}{MS: #1}}}
\newcommand{\edc}[1]{{\bf \textcolor{magenta}{EDC: #1}}}
\newcommand{\gs}[1]{{\bf \textcolor{red}{gs: #1}}}

\else
\newcommand{\XXX}[2]{}
\newcommand{\XXXR}[2]{}
\newcommand{\jbnote}[1]{}
\newcommand{\msnote}[1]{}
\newcommand{\edc}[1]{}
\newcommand{\gs}[1]{}

\fi

\newcommand{\descr}[1]{\smallskip\noindent\textbf{#1}}

\renewenvironment{thebibliography}[1]{
  \begin{oldthebibliography}{#1}
    \setlength{\itemsep}{0.1em}
    \setlength{\parskip}{0.1em}
}
{
  \end{oldthebibliography}
}

\renewcommand{\footnoterule}{%
  \kern -3pt
  \hrule width 1in 
  \kern 2pt
}

\usepackage{url}
\makeatletter
\def\url@leostyle{%
  \@ifundefined{selectfont}{\def\UrlFont{}}%
  {\def\UrlFont{}}%
}
\makeatother
\definecolor{darkred}{RGB}{153,0,0}
\definecolor{darkblue}{RGB}{0,0,119}
\hypersetup{colorlinks=true, linkcolor = darkred,   citecolor = darkred, urlcolor = black}

\usepackage{etoolbox}
\makeatletter
\patchcmd\@combinedblfloats{\box\@outputbox}{\unvbox\@outputbox}{}{%
   \errmessage{\noexpand\@combinedblfloats could not be patched}%
}%
 \makeatother
\AtBeginShipout{%
  \ifnum\value{page}>1 %
    \typeout{* Additional boxing of page `\thepage'}%
    \setbox\AtBeginShipoutBox=\hbox{\copy\AtBeginShipoutBox}%
  \fi
}

\begin{document}

\title{\bf The Web of False Information: Rumors, Fake News, Hoaxes, Clickbait, and Various Other Shenanigans}
\author{Savvas Zannettou$^{\star}$, Michael Sirivianos$^{\star}$, Jeremy Blackburn$^\ddagger$, Nicolas Kourtellis$^\dagger$\\[0.5ex]
\normalsize $^{\star}$Cyprus University of Technology, ${^\ddagger}$University of Alabama at Birmingham, $^\dagger$Telefonica Research \\
\normalsize sa.zannettou@edu.cut.ac.cy, michael.sirivianos@cut.ac.cy, blackburn@uab.edu, nicolas.kourtellis@telefonica.com }

\date{}

\maketitle

\begin{abstract}
A new era of Information Warfare has arrived.
Various actors, including state-sponsored ones, are weaponizing information on Online Social Networks to run false information campaigns with targeted manipulation of public opinion on specific topics.
These false information campaigns can have dire consequences to the public: mutating their opinions and actions, especially with respect to critical world events like major elections.
Evidently, the problem of false information on the Web is a crucial one, and needs increased public awareness, as well as immediate attention from law enforcement agencies, public institutions, and in particular, the research community.

In this paper, we make a step in this direction by providing a typology of the Web's false information ecosystem, comprising various types of false information, actors, and their motives.
We report a comprehensive overview of existing research on the false information ecosystem by identifying several lines of work: 1)~how the public perceives false information; 2)~understanding the propagation of false information; 3)~detecting and containing false information on the Web; and 4)~false information on the political stage.
In this work, we pay particular attention to political false information as: 1)~it can have dire consequences to the community (e.g., when election results are mutated) and 2)~previous work show that this type of false information propagates faster and further when compared to other types of false information. 
Finally, for each of these lines of work, we report several future research directions that can help us better understand and mitigate the emerging problem of false information dissemination on the Web. 

\end{abstract}

\section{Introduction}
Online Social Networks (OSNs) play an important role in the way people communicate and consume information. 
This is mainly because OSNs provide an ideal environment for communication and information acquisition, as users have access to a staggering amount of posts and articles that can share with others in real-time.
Unfortunately, OSNs have also become the mechanism for massive campaigns to diffuse false information~\cite{twitter_fake_news_problem_qz, facebook_fake_news_us_elections}.
In particular, recent reporting has highlighted how OSNs are exploited by powerful actors, potentially even state level, in order to manipulate individuals via targetted disinformation campaigns~\cite{facebook_russians_elections, nytimes_cambridge_analytica}.

The extensive dissemination of false information in OSNs can pose a major problem, affecting society in extremely worrisome ways. 
For example, false information can hurt the image of a candidate, potentially altering the outcome of an election.
During crisis situations (e.g., terrorist attacks, earthquakes, etc.), false information can cause wide spread panic and general chaos (e.g.,~\cite{panic_fake_news}).
False information diffusion in OSNs is achieved via diverse types of users, which typically have various motives. 
A diverse set of users are involved in the diffusion of false information, some unwittingly, and some with particular motives.
For example, terrorist organizations exploit OSNs to deliberately diffuse false information for propaganda purposes~\cite{al2015examining}.
Malicious users might utilize sophisticated automation tools (i.e., bots) or fake accounts that target specific benign users with the goal of influencing ideology.
No matter the motivation, however, the effects false information has on society clearly indicate the need for better understanding, measurement, and mitigation of false information in OSNs.

In this work, we provide a typology of the false information ecosystem that sheds light on the following questions: 1)~What are the various types and instances of false information on the Web? 2)~Who are the different actors that diffuse false information on the Web? and 3)~What are the motives behind the spread of false information?
Our typology is built after an extensive study of the existing literature, where we identify the following lines of work regarding the false information ecosystem:
\begin{itemize}
\item \textbf{User perception of false information.} This refers to how users perceive and interact with false information that is disseminated in OSNs.
For example, can users distinguish real stories from fake? If yes, what cues do they use to make this distinction?
\item \textbf{Propagation dynamics of false information in OSNs.}
Understanding the underlying propagation dynamics provides useful insights regarding the problem that can assist in detecting and mitigating false information.
In this line of work, we will review studies that focus on the propagation aspect of false information without expanding to detection and containment techniques.
\item \textbf{Detection and containment of false information.} Detecting and containing the propagation of false information is a desired outcome.
However, no robust platform, service or system is in-place that can effectively and efficiently mitigate the problem in a timely manner.
In this line of work, we will provide an overview of the community's efforts for detecting and containing the spread of false information on the Web.
\item \textbf{False information in politics.}
This line of work refers to work that focus on politics-related false information. Note, that this line of work overlaps with all of the other lines of work; however, we elect to devote a separate line of work for this survey for several reasons.
First, anecdotal evidence and existing literature suggest that false information is often disseminated for politics-related reasons.
Second, it can affect the course of history, especially in election periods.
For instance, during the 2016 US Election, as well as the 2017 French Elections, there were examples where trolls from 4chan tried to change the election outcome by disseminating false information about the candidate they opposed~\cite{4chan_disruptions_politics, 4chan_disruptions_politics2}.
On top of this, extensive anecdotal evidence suggests the involvement of state-sponsored troll accounts that actively try to mislead and mutate the opinions of users on OSNs~\cite{independent,zannettou2018disinformation}.
Also, previous work has shown that politics-related false information have a more intensive effect on social media when compared to other types of false information~\cite{vosoughi2018spread}.
Last, some of the previous work is specific to political context and is outside of the scope of the other three lines of work.
\end{itemize}
For each of these lines of work, we provide an overview of the most relevant research papers, as well as possible future research directions that will address existing gaps and will better help the community to alleviate the emerging problem of false information on the Web.
Note that the main focus of our literature review is towards computational approaches in understanding and tackling the problem of false information on the Web; however, we also report some previous work that sheds light on some other interesting features like interface design, socio-technical aspects, as well as systems that can help mitigate the problem.

\noindent \textbf{Contributions.} In summary, with the present study we make the following contributions: 
First, we provide a general overview of the false information ecosystem by proposing a typology. 
This will help other researchers that want to work on this field and have no previous experience with the topic.
Second, we provide a comprehensive overview of the related work that fall in one of the identified lines of work. 
This overview can act as an index for other researchers, which will be able to quickly identify previous work on a thematic, what methodology was used, and what Web communities previous work considered.
Furthermore, it can assist other researchers to quickly identify gaps in the literature.
Note that we acknowledge that the information ecosystem is huge and complex, hence we elect to have a broad scope in our survey, while paying particular attention to the political aspect of false information. 
This is mainly because it can have alarming consequences on the community and because previous work shows that the effects of false information are more intense in political contexts when compared to other types~\cite{vosoughi2018spread}.
Finally, for each identified line of work, we provide some future directions that the community can follow to extend the state-of-art on the Web's information ecosystem.

\noindent \textbf{Paper Organization.} The remainder of this paper is structured as follows: In Section~\ref{sec:taxonomy} we describe a typology for the false information ecosystem. Section~\ref{sec:user_perception} reports existing work on studying how user perceive and interact with false information on the Web, while Section~\ref{sec:propagation} describes the studies done on the propagation of information.
Studies about detecting and containing false information on the Web are presented in Section~\ref{sec:detection_containment}.
Section~\ref{sec:political} is devoted on false information in the politics stage whereas Section~\ref{sec:other} reports other relevant work that does not fit in any of the other lines of work.
Finally, we conclude in Section~\ref{sec:conclusions}.

\section{False Information Ecosystem Typology}
\label{sec:taxonomy}

In this section we present our typology, which we believe it will provide a succinct roadmap for future work.
The typology is based on~\cite{taxonomy_firstdraftnews} and extended to build upon the existing literature. 
Specifically, we describe the various types of false information that can be found in OSNs (Section~\ref{subsec:types}), the various types of actors that contribute in the distribution of false information (Section~\ref{subsec:actors}), as well as their motives (Section~\ref{subsec:motives}).
Note that our typology is different from concurrent work by Kumar and Shah~\cite{kumar2018false} as we provide a fine-grained distinction for the types of false information, the actors, and their motives.
Also, note that we make a best effort to cover as many aspects of the false information as per our knowledge; however, the typology should not be treated as an exhaustive representation of the false information ecosystem.

\subsection{Types of False Information} \label{subsec:types}
False information on the Web can be found in various forms, hence we propose the categorization of false information into eight types as listed below:
	\begin{itemize}
		 \item \textbf{Fabricated (F)~\cite{rubin2015towards}.} Completely fictional stories disconnected entirely from real facts. 
   This type is not new and it exists since the birth of journalism. 
    Some popular examples include fabricated stories about politicians and aliens~\cite{hillary_fake_news} (e.g., the story that Hillary Clinton adopted an alien baby).

    \item \textbf{Propaganda (P)~\cite{jowett2014propaganda}.} This is a special instance of the fabricated stories that aim to harm the interests of a particular party and usually has a political context.
This kind of false news is not new, as it was widely used during World War II and the Cold War.
Propaganda stories are profoundly utilized in political contexts to mislead people with the overarching goal of inflicting damage to a particular political party or nation-state.
Due to this, propaganda is a consequential type of false information as it can change the course of human history (e.g., by changing the outcome of an election).
Some recent examples of propaganda include stories about the Syria air strikes in 2018 or about specific movements like the BlackLivesMatter (see~\cite{medium_propaganda} for more examples).
            \item \textbf{Conspiracy Theories (CT)~\cite{fenster1999conspiracy}.} Refer to stories that try to explain a situation or an event by invoking a conspiracy without proof.
    Usually, such stories are about illegal acts that are carried out by governments or powerful individuals.
    They also typically present unsourced information as fact or dispense entirely with an ``evidence'' based approach, relying on leaps of faith instead.
    Popular recent examples of conspiracy theories include the Pizzagate theory (i.e., Clinton's campaign running a pedophile ring)~\cite{pizzagate} and conspiracies around the murder of Seth Rich~\cite{seth_rich} (e.g., Seth Rich was involved in the DNC email leaks).
	    \item \textbf{Hoaxes (H)~\cite{kumar2016disinformation}.} News stories that contain facts that are either false or inaccurate and are presented as legitimate facts. This category is also known in the research community either as half-truth~\cite{half-truth} or factoid~\cite{factoid} stories.
	    Popular examples of hoaxes are stories that report the false death of celebrities (e.g., the Adam Sadler death hoax~\cite{snopes_hoax_sadler}).
	    \item \textbf{Biased or one-sided (B).} Refers to stories that are extremely one-sided or biased.
	    In the political context, this type is known as Hyperpartisan news~\cite{potthast2017stylometric} and are stories that are extremely biased towards a person/party/situation/event.
    Some examples include the wide spread diffusion of false information to the alt-right community from small fringe Web communities like 4chan's /pol/ board~\cite{hine2017kek} and Gab, an alt-right echo chamber~\cite{zannettou2018gab}.
    \item \textbf{Rumors (R)~\cite{peterson1951rumor}.} Refers to stories whose truthfulness is ambiguous or never confirmed. 
    This kind of false information is widely propagated on OSNs, hence several studies have analyzed this type of false information.
    Some examples of rumors include stories around the 2013 Boston Marathon Bombings like the story that the suspects became citizens on 9/11 or that a Sandy Hook child was killed during the incident~\cite{boston_bombing_rumors}.
    \item \textbf{Clickbait (CL)~\cite{chen2015misleading}.} Refers to the deliberate use of misleading headlines and thumbnails of content on the Web. 
    This type is not new as it appeared years before, during the ``newspaper era,'' a phenomenon known as yellow journalism~\cite{campbell2001yellow}.
    However, with the proliferation of OSNs, this problem is rapidly growing, as many users add misleading descriptors to their content with the goal of increasing their traffic for profit or popularity~\cite{politifact_clickbait_profit}. 
    This is one of the least severe types of false information because if a user reads/views the whole content then he can distinguish if the headline and/or the thumbnail was misleading.
    \item \textbf{Satire News (S)~\cite{burfoot2009automatic}.} Stories that contain a lot of irony and humor. 
    This kind of news is getting considerable attention on the Web in the past few years. 
    Some popular examples of sites that post satire news are TheOnion~\cite{theonion} and SatireWire~\cite{satirewire}.
Usually, these sites disclose their satyric nature in one of their pages (i.e., About page). 
However, as their articles are usually disseminated via social networks, this fact is obfuscated, overlooked, or ignored by users who often take them at face value with no additional verification.
\end{itemize}

It is extremely important to highlight that there is an overlap in the aforementioned types of false information, thus it is possible to observe false information that may fall within multiple categories.
Here, we list two indicative examples to better understand possible overlaps: 1) a rumor may also use clickbait techniques to increase the audience that will read the story; and 2) propaganda stories, which are a special instance of a fabricated story, may also be biased towards a particular party.
These examples highlight that the false information ecosystem is extremely complex and the various types of false information need to be considered to mitigate the problem.

\subsection{False Information Actors} \label{subsec:actors}

In this section, we describe the different types of actors that constitute the false information propagation ecosystem.
We identified a handful of different actors that we describe below. %
\begin{itemize}
    \item \textbf{Bots \cite{boshmaf2011socialbot}.} In the context of false information, bots are programs that are part of a bot network (Botnet) and are responsible for controlling the online activity of several fake accounts with the aim of disseminating false information. Botnets are usually tied to a large number of fake accounts that are used to propagate false information in the wild. A Botnet is usually employed for profit by 3rd party organizations to diffuse false information for various motives (see Section \ref{motives_subsection} for more information on their possible motives).
    Note that various types of bots exist, which have varying capabilities; for instance, some bots only repost content, promote content (e.g., via vote manipulation on Reddit or similar platforms), and others post ``original'' content. However, this distinction is outside of the scope of this work, which provides a general overview of the information ecosystem on the Web.
    \item \textbf{Criminal/Terrorist Organizations \cite{al2015examining}.} Criminal gangs and terrorist organizations are exploiting OSNs as the means to diffuse false information to achieve their goals.
     A recent example is the ISIS terrorist organization that diffuses false information in OSNs for propaganda purposes~\cite{al2015examining}. 
    Specifically, they widely diffuse ideologically passionate messages for recruitment purposes.
    This creates an extremely dangerous situation for the community as there are several examples of individuals from European countries recruited by ISIS that ended-up perpetrating terrorist acts.
    \item \textbf{Activist or Political Organizations.} Various organizations share false information in order to either promote their organization, demote other rival organizations, or for pushing a specific narrative to the public. 
    A recent example include the National Rifle Association, a non-profit organiozation that advocates gun rights, which disseminated false information to manipulate people about guns~\cite{nra_fake_news}.
    Other examples include political parties that share false information, especially near major elections~\cite{allcott2017social}.
    \item \textbf{Governments~\cite{goverments_fake_news}.} Historically, governments were involved in the dissemination of false information for various reasons. More recently, with the proliferation of the Internet, governments utilize the social media to manipulate public opinion on specific topics. 
    Furthermore, there are reports that foreign goverments share false information on other countries in order to manipulate public opinion on specific topics that regard the particular country. Some examples, include the alleged involvement of the Russian government in the 2016 US elections~\cite{us_elections_russians} and Brexit referendum~\cite{brexit_russians}.
    \item \textbf{Hidden Paid Posters \cite{chen2013battling} and State-sponsored Trolls~\cite{zannettou2018disinformation}.} They are a special group of users that are paid in order to disseminate false information on a particular content or targeting a specific demographic.
    Usually, they are employed for pushing an agenda; e.g., to influence people to adopt certain social or business trends. 
    Similar to bots, these actors disseminate false information for profit.
    However, this type is substantially harder to distinguish than bots because they exhibit characteristics similar to regular users.
    \item \textbf{Journalists \cite{lee2004lying}.} Individuals that are the primary entities responsible for disseminating information both to the online and to the offline world.
    However, in many cases, journalists are found in the center of controversy as they post false information for various reasons.
    For example, they might change some stories so that they are more appealing, in order to increase the popularity of their platform, site, or newspaper.
    \item \textbf{Useful Idiots~\cite{useful_idiot}.} The term originates from the early 1950s in the USA as a reference to a particular political party's members that were manipulated by Russia in order to weaken the USA. 
    Useful idiots are users that share false information mainly because they are manipulated by the leaders of some organization or because they are naive.
    Usually, useful idiots are normal users that are not fully aware of the goals of the organization, hence it is extremely difficult to identify them.
    Like hidden paid posters, useful idiots are hard to distinguish and there is no study that focuses on this task. 
    
    \item \textbf{``True Believers'' and Conspiracy Theorists.} Refer to individuals that share false information because they actually believe that they are sharing the truth and that other people need to know about it. 
    For instace, a popular example is Alex Jones, which is a popular conspiracy theorist that shared false information about the Sandy Hook shooting~\cite{sandy_hook_wiki}.
    
    \item \textbf{Individuals that benefit from false information.} Refer to various individuals that will have a personal gain by disseminating false information. This is a very broad category ranging from common persons like an owner of a cafeteria to popular individuals like political persons.
    \item \textbf{Trolls \cite{mihaylov2015finding}.} The term troll is used in great extend by the Web community and refers to users that aim to do things to annoy or disrupt other users, usually for their own personal amusement. 
    An example of their arsenal is posting provocative or off-topic messages in order to disrupt the normal operation or flow of discussion of a website and its users.
    In the context of false information propagation, we define trolls as users that post controversial information in order to provoke other users or inflict emotional pressure.
    Traditionally, these actors use fringe communities like Reddit and 4chan to orcherstrate organized operations for disseminating false information to mainstream communities like Twitter, Facebook, and YouTube~\cite{zannettou2017web, hine2017kek}.
    \end{itemize}
    
Similarly to the types of false information, overlap may exist in actors too. 
Some examples include: 
1) Bots can be exploited by criminal organizations or political persons to disseminate false information~\cite{isis_bots}; and
2) Hidden paid posters and state-sponsored trolls can be exploited by political persons or organizations to push false information for a particular agenda~\cite{facebook_russians_elections}.

\subsection{Motives behind false information propagation} \label{subsec:motives}
\label{motives_subsection}

False information actors and types have different motives behind them. 
Below we describe the categorization of motives that we distinguish:
\begin{itemize}
    \item \textbf{Malicious Intent.} 
    Refers to a wide spectrum of intents that drive actors that want to hurt others in various ways.
    Some examples include inflicting damage to the public image of a specific person, organization, or entity. 
    \item \textbf{Influence.} 
    This motive refers to the intent of misleading other people in order to influence their  decisions, or manipulate public opinion with respect to specific topics. 
    This motive can be distinguished into two general categories; 1)~aiming to get leverage or followers (\emph{power}) and 2)changing the norms of the public by disseminating false information.
    This is particularly worrisome on political matters~\cite{fake_news_politics}, where individuals share false information to enhance an individuals' public image or to hurt the public image of opposing politicians, especially during election periods.
    \item \textbf{Sow Discord.} In specific time periods individuals or organizations share false information to sow confusion or discord to the public. 
    Such practices can assist in pushing a particular entity's agenda; we have seen some examples on the political stage where foreign governments try to seed confusion in another country's public for their own agenda~\cite{russia_discord}.
    \item \textbf{Profit.}
    Many actors in the false information ecosystem seek popularity and monetary profit for their organization or website. 
    To achieve this, they usually disseminate false information that increases the traffic on their website. 
    This leads to increased ad revenue that results in monetary profit for the organization or website, at the expense of manipulated users.
    Some examples include the use of clickbait techniques, as well as fabricated news to increase views of articles from fake news sites that are disseminated via OSNs~\cite{politifact_clickbait_profit, fake_news_profit}
    \item \textbf{Passion.} A considerable amount of users are passionate about a specific idea, organization, or entity.
    This affects their judgment and can contribute to the dissemination of false information.
    Specifically, passionate users are blinded by their ideology and perceive the false information as correct, and contribute in its overall propagation~\cite{fake_news_passion}.
    \item \textbf{Fun.} As discussed in the previous section, online trolls are usually diffusing false information for their amusement.
  Their actions can sometimes inflict considerable damage to other individuals (e.g., see Doxing~\cite{snyder2017fifteen}), and thus should not be taken lightly.
\end{itemize}

Again, similarly to Sections~\ref{subsec:types} and \ref{subsec:actors}, we have overlap among the presented motives. 
For instance, a political person may disseminate false information for political influence and because he is passionate about a specific idea.

\section{User Perception of False Information}
\label{sec:user_perception}

In this section, we describe work that study how users perceive and interact with false information on OSNs.
Existing work use the following methodologies in understanding how false information is perceived by users:
(i) by analyzing large-scale datasets obtained from OSNs; and
(ii) by receiving input from users either from questionnaires, interviews, or through crowdsourcing marketplaces (e.g., Amazon Mechanical Turk, AMT~\cite{mechanical_turk}).
Table~\ref{tbl:user_perception_papers_summary} summarizes the studies on user perception, as well as their methodology and the considered OSN.
Furthermore, we annotate each entry in Table~\ref{tbl:user_perception_papers_summary} with the type of false information that each work considers.
The remainder of this section provides an overview of the studies on understanding users' perceptions on false information.

\begin{table}[]
\centering
\resizebox{.8\columnwidth}{!}{
\begin{tabular}{@{}cccc@{}}
\toprule
\textbf{Platform} & \textbf{OSN data analysis}                                                                                                                                                                                                                                                     & \textbf{Questionnaires/Interviews}                                                                                                                                    & \textbf{Crowdsourcing platforms}                \\ \midrule
Twitter           & \begin{tabular}[c]{@{}c@{}}Kwon et al.~\cite{kwon2013aspects} \textbf{(R)},\\ Zubiaga et al.~\cite{zubiaga2016analysing} \textbf{(R)},\\
Thomson et al.~\cite{thomson2012trusting} \textbf{(R)}\end{tabular}                                                                                                                                         & Morris et al.~\cite{morris2012tweeting} \textbf{(CA)}
                                                                                                                                                                    & \begin{tabular}[c]{@{}c@{}}Ozturk et al.~\cite{ozturk2015combating}~\textbf{(R)},\\McCreadie et al.~\cite{mccreadie2015crowdsourced}~\textbf{(R)}\end{tabular}\\ \midrule 

Facebook          & \begin{tabular}[c]{@{}c@{}}Zollo et al.~\cite{zollo2015emotional}~\textbf{(CT)},\\ Zollo et al.~\cite{zollo2015debunking}~\textbf{(CT)},\\ Bessi et al.~\cite{bessi2015science}~\textbf{(CT)}\end{tabular} & Marchi~\cite{marchi2012facebook} \textbf{(B)}                                                                                                                               & X                                  \\ \midrule
Other               & Dang et al.~\cite{dang2016toward} \textbf{(R)}                                                                                                                                                                                                                                                                                                                                                                                                                                                                                                                     & \begin{tabular}[c]{@{}c@{}}Chen et al.~\cite{chen2015why}~\textbf{(F)},\\ Kim and Bock~\cite{kim2011study}~\textbf{(R)},\\ Feldman~\cite{feldman2011partisan}~\textbf{(B)},\\
Brewer et al.~\cite{brewer2013impact}~\textbf{(S)}\\
Winerburg and McGrew~\cite{wineburg2017lateral} \textbf{(CA)}\end{tabular} & X                                               \\ \bottomrule
\end{tabular}
}
\caption{Studies of user perception and interaction with false information on OSNs. The table depicts the main methodology of each paper as well as the considered OSN (if any). Also, where applicable, we report the type of false information that is considered (see bold markers and cf. with Section~\ref{subsec:types}).
}
\label{tbl:user_perception_papers_summary}
\end{table}

\subsection{OSN data analysis}
Previous work focuses on extracting meaningful insights by analyzing data obtained from OSNs.
From Table~\ref{tbl:user_perception_papers_summary} we observe that previous work, leverages data analysis techniques to mainly study how users perceive and interact with rumors and conspiracy theories.

\descr{Rumors.} Kwon et al.~\cite{kwon2013aspects} study the propagation of rumors in Twitter, while considering findings from social and psychological studies. 
By analyzing 1.7B tweets, obtained from~\cite{cha2010measuring}, they find that: 
1) users that spread rumors and non-rumors have similar registration age and number of followers; 
2) rumors have a clearly different writing style; 
3) sentiment in news depends on the topic and not on the credibility of the post; and 
4) words related to social relationships are more frequently used in rumors.
Zubiaga et al.~\cite{zubiaga2016analysing} analyze 4k tweets related to rumors by using journalists to annotate rumors in real time.
Their findings indicate that true rumors resolved faster than false rumors and that the general tendency for users is to support every unverified rumor. 
However, the latter is less prevalent to reputable user accounts (e.g., reputable news outlets) that usually share information with evidence.
Thomson et al.~\cite{thomson2012trusting} study Twitter's activity regarding the Fukushima Daiichi nuclear power plant disaster in Japan.
The authors undertake a categorization of the messages according to their user, location, language, type, and credibility of the source.
They observe that anonymous users, as well as users that live far away from the disaster share more information from less credible sources.
Finally, Dang et al.~\cite{dang2016toward} focus on the users that interact with rumors on Reddit by studying a popular false rumor (i.e., Obama is a Muslim). 
Specifically, they distinguish users into three main categories: the ones that support false rumors, the ones that refute false rumors and the ones that joke about a rumor.
To identify these users they built a Naive Bayes classifier that achieves an accuracy of 80\% and find that more than half of the users joked about this rumor, 25\% refuted the joke and only 5\% supported this rumor.

\descr{Conspiracy Theories.} Zollo et al.~\cite{zollo2015emotional} study the emotional dynamics around conversations regarding science and conspiracy theories. 
They do so by collecting posts from 280k users on Facebook pages that post either science or conspiracy theories posts. 
Subsequently, they use Support Vector Machines (SVMs) to identify the sentiment values of the posts, finding that sentiment is more negative on pages with conspiracy theories.
Furthermore, they report that as conversations grow larger, the overall negative sentiment in the comments increases.
In another work, Zollo et al.~\cite{zollo2015debunking} perform a quantitative analysis of 54M Facebook users. finding the existence of well-formed communities for the users that interact with science and conspiracy news. 
They note that users of each community interact within the community and rarely outside of it. 
Also, debunking posts are rather inefficient and user exposure to such content increases the overall interest in conspiracy theory posts.
Similarly, Bessi et al.~\cite{bessi2015science} study how conspiracy theories and news articles are consumed on Facebook, finding that polarized users contribute more in the diffusion of conspiracy theories, whereas this does not apply for news and their respective polarized users.

\subsection{Questionnaires/Interviews}
To get insights on how users perceive the various types of false information, some of the previous work conducted questionnaires or interviews. 
The majority of the work aims to understand how younger users (students or teenagers) interact and perceive false information.

\descr{Credibility Assessment.} Morris et al.~\cite{morris2012tweeting} highlight that users are influenced by several features related to the author of a tweet like their Twitter username when assessing the credibility of information. 
Winerburg and McGrew~\cite{wineburg2017lateral} study whether users with different backgrounds have differences in their credibility assessments. 
To achieve this they conducted experiments with historians, fact-checkers, and undergraduate students, finding that historians and students can easily get manipulated by official-looking logos and domain names.

\descr{Biased.} Marchi~\cite{marchi2012facebook} focus on how teenagers interact with news on Facebook by conducting interviews with 61 racially diverse teenagers.
The main findings of this study is that teenagers are not very interested in consuming news (despite the fact that their parents do) and that they demonstrate a preference to news that are opinionated when compared to objective news.
Similarly, Feldman~\cite{feldman2011partisan} focus on biased news and conduct 3 different studies with the participants randomly exposed to 2 biased and 1 non-biased news.
The participants were asked to provide information about the news that allowed the authors to understand the perceived bias.
They find that participants are capable of distinguishing bias in news articles; however, participants perceived lower bias in news that agree with their ideology/viewpoints.

\descr{Fabricated.} Chen et al.~\cite{chen2015why} use questionnaires on students from Singapore with the goal to unveil the reasons that users with no malicious intent share false information on OSNs. 
They highlight that female students are more prone in sharing false information, and that students are willing to share information of any credibility just to initiate conversations or because the content seems interesting.

\descr{Rumors.} Kim and Bock~\cite{kim2011study} study the rumor spreading behavior in OSNs from a psychological point of view by undertaking questionnaires on Korean students. 
They find that users' beliefs results in either positive or negative emotion for the rumor, which affects the attitude and behavior of the users towards the rumor spreading.

\descr{Satire.} Brewer et al.~\cite{brewer2013impact} indicate that satirical news programs can affect users' opinion and political trust, while at the same time users tend to have stronger opinion on matters that they have previously seen in satirical programs.

\subsection{Crowdsourcing platforms}
Other related work leverages crowdsourcing platform to get feedback from users about false information. We note that, to the best of our knowledge, previous work that used crowdsourcing platforms focused on rumors.

\descr{Rumors.} Ozturk et al.~\cite{ozturk2015combating} study how users perceive health-related rumors and if their are willing to share them on Twitter.
For acquiring the rumors, they crawl known health-related websites such as Discovery, Food Networks and National Institute of Health websites. 
To study the user perceptions regarding these rumors, they use AMT where they query 259 participants about ten handpicked health-related rumors.
The participants were asked whether they will share a specific rumor or a message that refutes a rumor or a rumor that had a warning on it (i.e., ``this message appeared in a rumor website''). 
Their results indicate that users are less likely to share a rumor that is accompanied with a warning or a message that refutes a rumor. 
Through simulations, they demonstrate that this approach can help in mitigating the spread of rumors on Twitter.
Finally, McCreadie et al.~\cite{mccreadie2015crowdsourced} use crowdsourcing on three Twitter datasets related to emergency situations during 2014, in order to record users' identification of rumors. 
Their results note that users were able to label most of the tweets correctly, while they note that tweets that contain controversial information are harder to distinguish.

\subsection{User Perception - Future Directions}
The studies discussed in this section aim to shed light on how users \emph{perceive} false information on the Web.
Overall the main take-away points from the reviewed related work are:
1) teenagers are not interested in consuming news;
2) students share information of any credibility just to initate conversations;
3) in most cases, adults can identify bias in news and this task is harder when the news are biased towards the reader's ideology;
and 4) users can mostly identify rumors except the ones that contain controversial information.
After reviewing the literature, we identify a few gaps in our understanding of false information perception.
First, there is a lack of rigorous temporal analysis of user perception around the dissemination of false information and/or conspiracy theories.
For example, perceptions might differ during the course of evolution of any particular conspiracy theory.
Next, none of the studies reviewed take into consideration the interplay between \emph{multiple} OSNs.
Users on one platform might perceive the same false information differently depending on a variety of factors.
For example, certain communities might be focused around a particular topic, affecting their susceptibility to false information on that topic, or the way the platform calls home presents information (e.g., news feed) can potentially influence how it users perceive false information.
This issue becomes further muddied when considering users that are active on multiple platforms.
In particular, we note that there is a substantial gap regarding \emph{which} OSNs have been studied with respect to false information; e.g., YouTube, which has become a key player in information consumption.
Finally, we would like to note that the European Union has recently approved a new regulation with regard to data protection, known as General Data Protection Regulation (GDPR)~\cite{gdpr}. 
Therefore, we strongly advise researchers working in this line of work to study the new regulation and make sure that users give their explicit consent for participating in studies that aim to understand how users perceive false information.

\section{Propagation of False Information}
\label{sec:propagation}

Understanding the dynamics of false information is of paramount importance as it gives useful insights regarding the problem. 
Table~\ref{tbl:propagation_summary} summarizes the studies of false information propagation at OSNs, their methodology, as well as the corresponding type of false information according to the typology in Section~\ref{subsec:types}. 
The research community focuses on studying the propagation by either employing data analysis techniques or mathematical and statistical approaches. 
Furthermore, we note the efforts done on providing systems that visualize the propagation dynamics of false information.  
Below, we describe the studies that are mentioned in Table~\ref{tbl:propagation_summary} by dedicating a subsection for each type of methodology.

\begin{table}[]
\centering
\resizebox{0.8\columnwidth}{!}{
\begin{tabular}{cccc}
\hline
\textbf{Platform} & \textbf{OSN data analysis}                                                                              & \textbf{Epidemic \& Statistical Modeling}                                              & \textbf{Systems} \\ \hline
Twitter           & \begin{tabular}[c]{@{}c@{}}Mendoza et al.~\cite{mendoza2010twitter}~\textbf{(R)},\\Oh et al.~\cite{oh2010exploration}~\textbf{(R)},\\
Andrews et al.~\cite{andrews2016keeping}~\textbf{(R)},\\
Gupta et al.~\cite{gupta20131}~\textbf{(F)},\\
Starbird et al.~\cite{starbird2014rumors}~\textbf{(R)},\\ 
Arif et al.~\cite{arif2016information}~\textbf{(R)},\\
Situngkir~\cite{situngkir2011spread}~\textbf{(H)},\\ 
Nadamoto et al.~\cite{nadamoto2013analysis}~\textbf{(R)},\\
Vosoughi et al.~\cite{vosoughi2018spread}~\textbf{(F)}\end{tabular}                                         & \begin{tabular}[c]{@{}c@{}}Jin et al.~\cite{jin2013epidemiological}~\textbf{(R)}, \\ Doerr et al.~\cite{doerr2012why}~\textbf{(R)},\\
Jin et al.~\cite{jin2014misinformation}~\textbf{(R)}\\\end{tabular}   
 & \begin{tabular}[c]{@{}c@{}}Finn et al.~\cite{finn2014investigating}~\textbf{(R)},\\ Shao el at.~\cite{shao2016hoaxy}~\textbf{(F)}\end{tabular}\\ \midrule
Facebook          & \begin{tabular}[c]{@{}c@{}}Friggeri et al.~\cite{friggeri2014rumor}~\textbf{(R)},\\ Del Vicario et al.~\cite{del2016spreading}~\textbf{(CT)},\\ Anagnostopoulos et al.~\cite{anagnostopoulos2014viral}~\textbf{(CT)}\end{tabular} & Bessi~\cite{bessi2017statistical}~\textbf{(CT)}                                                                                    & X                \\ \midrule
Other            & \begin{tabular}[c]{@{}c@{}}Ma and Li~\cite{ma2016rumor}~\textbf{(R)},\\
Zannettou et al.~\cite{zannettou2017web}~\textbf{(B)}                                                                                             \end{tabular}
 & \begin{tabular}[c]{@{}c@{}}Shah et al.~\cite{shah2011rumors}~\textbf{(R)}, \\ Seo et al.~\cite{seo2012identifying}~\textbf{(R)} ,\\ Wang et al.~\cite{wang2014rumor}~\textbf{(R)}\end{tabular} &    Dang et al.~\cite{dang2016what}~\textbf{(R)}             \\ \midrule
Sina Weibo        & X & Nguyen et al.~\cite{nguyen2012sources}~\textbf{(R)}                                                                                    & X                \\ \hline
\end{tabular}
}
\caption{Studies the focus on the propagation of false information on OSNs. The table summarizes the main methodology of each paper as well as the considered OSNs. Also, we report the type of false information that is considered (see bold markers and cf. with Section~\ref{subsec:types}
}
\label{tbl:propagation_summary}
\end{table}

\subsection{OSN Data Analysis}

\descr{Rumors.} Mendoza et al.~\cite{mendoza2010twitter} study the dissemination of false rumors and confirmed news on Twitter the days following the 2010 earthquake in Chile. 
They analyze the propagation of tweets for confirmed news and for rumors finding that the propagation of rumors differs from the confirmed news and that an aggregate analysis on the tweets can distinguish the rumors from the confirmed news.
Similarly, Starbird et al.~\cite{starbird2014rumors} study rumors regarding the 2013 Boston Bombings on Twitter and confirm both findings from Mendoza et al.~\cite{mendoza2010twitter}. %
In a similar notion, Nadamoto et al.~\cite{nadamoto2013analysis} analyze the behavior of the Twitter community during disasters (Great East Japan Earthquake in 2011) when compared to a normal time period; finding that the spread of rumors during a disaster situation is different from a normal situation. That is in disaster situations, the hierarchy of tweets is shallow whereas in normal situations the tweets follow a deep hierarchy. 

Others focused on understanding how rumors can be controlled and shed light on which types of accounts can help stop the rumor spread.
Oh et al.~\cite{oh2010exploration} study Twitter data about the 2010 Haiti Earthquake and find that credible sources contribute in rumor controlling, while Andrews et al.~\cite{andrews2016keeping} find that official accounts can contribute in stopping the rumor propagation by actively engaging in conversations related to the rumors.

Arif et al.~\cite{arif2016information} focus on the 2014 hostage crisis in Sydney. 
Their analysis include three main perspectives; (i) volume (i.e., number of rumor-related messages per time interval); (ii) exposure (i.e., number of individuals that were exposed to the rumor) and (iii) content production (i.e., if the content is written by the particular user or if it is a share).
Their results highlight all three  perspectives are important in understanding the dynamics of rumor propagation.
Friggeri et al.~\cite{friggeri2014rumor} use known rumors that are obtained through Snopes~\cite{snopes}, a popular site that covers rumors, to study the propagation of rumors on Facebook.
Their analysis indicates that rumors' popularity is bursty and that a lot of rumors change over time, thus creating rumor variants.
These variants aim to reach a higher popularity burst.
Also, they note that rumors re-shares which had a comment containing a link to Snopes had a higher probability to be deleted by their users. 

Finally, Ma and Li~\cite{ma2016rumor} study the rumor propagation process when considering a two-layer network; one layer is online (e.g., Twitter) and one layer is offline (e.g., face-to-face). 
Their simulations indicate that rumor spread is more prevalent in a two-layer network when compared with a single-layer offline network. 
The intuition is that in an offline network the spread is limited by the distance, whereas this constraint is eliminated in a two-layer network that has an online social network. 
Their evaluation indicates that in a two-layer network the spreading process on one layer does not affect the spreading process of the other layer; mainly because the interlayer transfer rate is less effective from an offline to an online network when compared with that from an OSN.

\descr{Fabricated.} Gupta et al.~\cite{gupta20131} study the propagation of false information on Twitter regarding the 2013 Boston Marathon Bombings. 
To do so, they collect 7.9M unique tweets by using keywords about the event.
Using real annotators, they annotate 6\% of the whole corpus that represents the 20 most popular tweets during this crisis situation (i.e., the 20 tweets that got retweeted most times).
Their analysis indicate that 29\% of the tweets were false and a large number of those tweets were disseminated by reputable accounts.
This finding contradicts with the findings of Oh et al.~\cite{oh2010exploration}, which showed that credible accounts help stop the spread of false information, hence highlighting that reputable accounts can share bad information too.
Furthermore, they note that out of the 32K accounts that were created during the crisis period, 19\% of them were deleted or suspended by Twitter, indicating that accounts were created for the whole purpose of disseminating false information.

Vosoughi et al.~\cite{vosoughi2018spread} study the diffusion of false and true stories in Twitter over the course of 11 years. 
They find that false stories propagate faster, farther, and more broadly when compared to true stories.
By comparing the types of false stories, they find that these effects were more intensive for political false stories when compared to other false stories (e.g., related to terrorism, science, urban legends, etc.).

\descr{Hoaxes.} Situngkir~\cite{situngkir2011spread} observe an empirical case in Indonesia to understand the spread of hoaxes on Twitter. 
Specifically, they focus on a case where a Twitter user with around 100 followers posted a question of whether a well-known individual is dead.
Interestingly, the hoax had a large population spread within 2 hours of the initial post and it could be much larger if a popular mainstream medium did not publicly deny the hoax.
Their findings indicate that a hoax can easily spread to the OSN if there is collaboration between the recipients of the hoax.
Again, this work highlights, similarly to Oh et al.~\cite{oh2010exploration} that reputable accounts can help in mitigating the spread of false information.

\descr{Conspiracy Theories.} Del Vicario et al.~\cite{del2016spreading} analyze the cascade dynamics of users on Facebook when they are exposed to conspiracy theories and scientific articles.
They analyze the content of 67 public pages on Facebook that disseminate conspiracy theories and science news.
Their analysis indicates the formulation of two polarized and homogeneous communities for each type of information.
Also, they note that despite the fact that both communities have similar content consumption patterns, they have different cascade dynamics.
Anagnostopoulos et al.~\cite{anagnostopoulos2014viral} study the role of homophily and polarization on the spread of false information by analyzing 1.2M Facebook users that interacted with science and conspiracy theories. 
Their findings indicate that user's interactions with the articles correlate with the interactions of their friends (homophily) and that frequent exposure to conspiracy theories (polarization) determines how viral the false information is in the OSN.

\descr{Biased.} Zannettou et al.~\cite{zannettou2017web}, motivated by the fact that the information ecosystem consists of multiple Web communities, study the propagation of news across multiple Web communities.
To achieve this, they study URLs from 99 mainstream and alternative news sources on three popular Web communities: Reddit, Twitter, and 4chan. 
Furthermore, they set out to measure the influence that each Web community has to each other, using a statistical model called Hawkes Processes.
Their findings indicate that small fringe communities within Reddit and 4chan have a substantial influence to mainstream OSNs like Twitter. 

\subsection{Epidemic and Statistical Modeling}
\descr{Rumors.} Jin et al.~\cite{jin2013epidemiological} use epidemiological models to characterize cascades of news and rumors in Twitter. Specifically, they use the SEIZ model~\cite{bettencourt2006power} which divides the user population in four different classes based on their status; 
(i) Susceptible; 
(ii) Exposed; 
(iii) Infected 
and (iv) Skeptic.
Their evaluation indicates that the SEIZ model is better than other models and it can be used to distinguish rumors from news in Twitter.
In their subsequent work, Jin et al.~\cite{jin2014misinformation} perform a quantitative analysis on Twitter during the Ebola crisis in 2014. By leveraging the SEIZ model, they show that rumors spread in Twitter the same way as legitimate news.

Doerr et al.~\cite{doerr2012why} use a mathematical approach to prove that rumors spread fast in OSNs (similar finding with Vosoughi et al.~\cite{vosoughi2018spread}).
For their simulations they used real networks that represent the Twitter and Orkut Social Networks topologies obtained from \cite{cha2010measuring} and SNAP~\cite{snap_datasets}, respectively. 
Intuitively, rumors spread fast because of the combinations of few large-degree nodes and a large number of small-degree nodes. 
That is, small-degree nodes learn a rumor once one of their adjacent nodes knows it, and then quickly forward the rumor to all adjacent nodes. 
Also, the propagation allows the diffusion of rumors between 2 large-degree nodes, thus the rapid spread of the rumor in the network.

Several related work focus on finding the source of the rumor.
Specifically, Shah et al.~\cite{shah2011rumors} focus on detecting the source of the rumor in a network by defining a new rumor spreading model and by forming the problem as a maximum likelihood estimation problem.
Furthermore, they introduce a new metric, called \textit{rumor centrality}, which essentially specifies the likelihood that a particular node is the source of the rumor. 
This metric is evaluated for all nodes in the network by using a simple linear time message-passing algorithm, hence the source of the rumor can be found by selecting the node with the highest rumor centrality.
In their evaluation, they used synthetic small-world and scale-free real networks to apply their rumor spreading model and they show that they can distinguish the source of a rumor with a maximum error of 7-hops for general networks, and with a maximum error of 4-hops for tree networks. 
Seo et al.~\cite{seo2012identifying} aim to tackle the same problem by injecting monitoring nodes on the social graph.
They propose an algorithm that considers the information received by the monitoring nodes to identify the source.
They indicate that with sufficient number of monitoring nodes they can recognize the source with high accuracy.
Wang et al.~\cite{wang2014rumor} aim to tackle the problem from a statistical point of view. 
They propose a new detection framework based on rumor centrality, which is able to support multiple snapshots of the network during the rumor spreading.
Their evaluation based on small-world and scale-free real networks note that by using two snapshots of the network, instead of one, can improve the source detection. 
Finally, Nguyen et al.~\cite{nguyen2012sources} aim to find the $k$ most suspected users where a rumor originates by proposing the use of a reverse diffusion process in conjunction with a ranking process.

\descr{Conspiracy Theories.} Bessi~\cite{bessi2017statistical} perform a statistical analysis of a large corpus (354k posts) of conspiracy theories obtained from Facebook pages. 
Their analysis is based on the Extreme Value Theory branch of statistics~\cite{extreme_value_theory} and they find that extremely viral posts (greater than 250k shares) follow a Poisson distribution. 

\subsection{Systems}

\descr{Rumors.} Finn et al.~\cite{finn2014investigating} propose a web-based tool, called TwitterTrails, which enables users to study the propagation of rumors in Twitter. 
TwitterTrails demonstrates indications for bursty activity, temporal characteristics of propagation, and visualizations of the re-tweet networks. 
Furthermore, it offers advanced metrics for rumors such as level of visibility and community's skepticism towards the rumor (based on the theory of h-index~\cite{h_index}).
Similarly, Dang et al.~\cite{dang2016what} propose RumourFlow, which visualizes rumors propagation by adopting modeling and visualization tools.
It encompasses various analytical tools like semantic analysis and similarity to assist the user in getting a holistic view of the rumor spreading and its various aspects. 
To demonstrate their system, they collect rumors from Snopes and conversations from Reddit.

\descr{Fabricated.} Shao et al.~\cite{shao2016hoaxy} propose Hoaxy, a platform that provides information about the dynamics of false information propagation on Twitter as well as the respective fact checking efforts.

\subsection{Propagation of False Information - Future Directions}
In this section, we provided an overview of the existing work that focuses on the propagation of false information on the Web. 
Some of the main take-aways from the literature review on the propagation of false information are:
1) Accounts on social networks are created with the sole purpose of disseminating false information;
2) False information is more persistent than corrections;
3) The popularity of false information follow a bursty activity;
4) Users on Web communities create polarized communities that disseminate false information;
5) Reputable or credible accounts are usually useful in stopping the spread of false information; however we need to pay particular attention as previous work (see Gupta et al.~\cite{gupta20131}) has showed that they also share false information;
6) Being able to detect the source of false information is a first step towards stopping the spread of false information on Web communities and several approaches exist that offer acceptable performance.
As future directions to this line of work, we propose studying the problem from a multi-platform point of view.
That is, study how information propagates across \emph{multiple communities} and fusing information that exists in \emph{multiple formats} (e.g., images, textual claims, URLs, video, etc.).
Furthermore, systems or tools that visualize the propagation of information across OSNs do not exist.
These type of tools will enable a better understanding of false information propagation, as well as finding the source of information.
Finally, to the best of our knowledge, the propagation of information via \emph{orchestrated campaigns} has not been rigorously studied by the research community.
An example of such a campaign is the posting comments in YouTube video by users of 4chan~\cite{hine2017kek}.

\section{Detection and Containment of False Information}
\label{sec:detection_containment}

\subsection{Detection of false information}
Detecting false information is not a straightforward task, as it appears in various forms, as discussed in Section~\ref{sec:taxonomy}.
Table~\ref{tbl:detection_summary} summarizes the studies that aim to solve the false information detection problem, as well as their considered OSNs and their methodology.
Most studies try to solve the problem using handcrafted features and conventional machine learning techniques.
Recently, to avoid using handcrafted features, the research community used neural networks to solve the problem (i.e., Deep Learning techniques).
Furthermore, we report some systems that aim to inform users about detected false information. 
Finally, we also note a variety of techniques that are proposed for the detection and containment of false information, such as epidemiological models, multivariate Hawkes processes, and clustering.
Below, we provide more details about existing work grouped by methodology and the type of information, according to Table~\ref{tbl:detection_summary}.

\begin{table}[t]
\centering
\resizebox{0.8\textwidth}{!}{
\begin{tabular}{@{}cccc@{}}
\toprule
\textbf{Platform}                                                          & \textbf{Machine Learning}                                                                                                                                                                                                                                              & \textbf{Systems}                                                                          & \textbf{Other models/algorithms}                                                                                                                       \\ \midrule
Twitter                                                                    & \begin{tabular}[c]{@{}c@{}}Castillo et al.~\cite{castillo2011information}~\textbf{(CA)},\\ Gupta and Kumaraguru~\cite{gupta2012credibility}~\textbf{(CA)},\\ Kwon et al.~\cite{kwon2013prominent}~\textbf{(R)},\\ Yang et al.~\cite{yang2015emerging}~\textbf{(R)},\\ Liu et al.~\cite{liu2015real}~\textbf{(R)},\\ Wu et al.~\cite{wu2017gleaning}~\textbf{(R)},\\ Gupta et al.~\cite{gupta2014tweetcred}~\textbf{(CA)}, \\ AlRubaian et al.~\cite{alrubaian2015multistage}~\textbf{(CA)},\\ Hamidian and Diab~\cite{hamidian2016rumor}~\textbf{(R)},\\ Giasemidis et al.~\cite{giasemidis2016determining}~\textbf{(R)} ,\\ Kwon et al.~\cite{kwon2017rumor}~\textbf{(R)},\\
Volkova et al.~\cite{volkova2017separating}~\textbf{(CA)}  \end{tabular} 
                    & \begin{tabular}[c]{@{}c@{}}Resnick et al.~\cite{resnick2014rumorlens}~\textbf{(R)},\\ Vosoughi et al.~\cite{vosoughi2015human}~\textbf{(R)},\\ Jaho et al.~\cite{jaho2014alethiometer}~\textbf{(CA)}\end{tabular} & \begin{tabular}[c]{@{}c@{}}Qazvinian et al.~\cite{qazvinian2011rumor}~\textbf{(R)} \\(rumor retrieval model),\\ Zhao el al.~\cite{zhao2015enquiring}~\textbf{(R)} \\(clustering),\\ Farajtabar et al.~\cite{farajtabar2017fake}~\textbf{(F)}\\(hawkes process),\\
Kumar and Geethakumari~\cite{kumar2014detecting}~\textbf{(F)}\\
(algorithm with psychological cues)\end{tabular}       \\ \midrule
Sina Weibo                                                                 & \begin{tabular}[c]{@{}c@{}}Yang et al.~\cite{yang2012automatic}~\textbf{(R)},\\ Wu et al.~\cite{wu2015false}~\textbf{(R)},\\ Liang et al.~\cite{liang2015rumor}~\textbf{(R)},\\ Zhang et al.~\cite{zhang2015automatic}~\textbf{(R)},\end{tabular}                                                                                                                                                                      & Zhou et al.~\cite{zhou2015real}~\textbf{(CA)}                                                                          & X                                                                                                                                                      \\ \midrule
\begin{tabular}[c]{@{}c@{}}Twitter and \\ Sina Weibo\end{tabular}          & \begin{tabular}[c]{@{}c@{}}Ma et al.~\cite{ma2015detect}~\textbf{(CA)}\\Ma et al.~\cite{ma2016detecting}~\textbf{(R)} \end{tabular}                                                                                                                                                                                                                                                                    & X                                                                                         & \begin{tabular}[c]{@{}c@{}}Jin et al.~\cite{jin2016news}~\textbf{(CA)}\\(graph optimization)\end{tabular}                                                                                                                          \\ \midrule
Facebook                                                                   & \begin{tabular}[c]{@{}c@{}}Tacchini et al.~\cite{tacchini2017some}~\textbf{(H)},\\ Conti et al.~\cite{conti2017s}~\textbf{(CT)}\end{tabular}                                                                                                                                                                                                                    & X                                                                                         & X                                                                                                                                                      \\ \midrule
\begin{tabular}[c]{@{}c@{}}Wikipedia and/or \\ other articles\end{tabular} & \begin{tabular}[c]{@{}c@{}}Qin et al.~\cite{qin2016spotting}~\textbf{(R)},\\ Rubin et al.~\cite{rubin2016fake}~\textbf{(S)},\\ Kumar et al.~\cite{kumar2016disinformation}~\textbf{(H)},\\ Chen et al.~\cite{chen2015misleading}~\textbf{(CL)},\\ Chakraborty et al.~\cite{chakraborty2016stop}~\textbf{(CL)},\\ Potthast et al.~\cite{potthast2016clickbait}~\textbf{(CL)},\\ Biyani et al.~\cite{Biyani2016}~\textbf{(CL)},\\Wang~\cite{wang2017liar}~\textbf{(F)},\\ Anand et al. ~\cite{anand2016we}~\textbf{(CL)}\end{tabular}                                                                 & X                                                                                         & \begin{tabular}[c]{@{}c@{}}Potthast et al.~\cite{potthast2017stylometric}~\textbf{(B)}\\(unmasking)\end{tabular}                                                                                                                            \\ \midrule
Other                                                                      & \begin{tabular}[c]{@{}c@{}}
Afroz et al.~\cite{afroz2012detecting}~\textbf{(H)},\\
Maigrot et al.~\cite{maigrot2016mediaeval}~\textbf{(H)},\\
Zannettou et al.~\cite{zannettou2018good}~\textbf{(CL)}\end{tabular}                                                                                                                                                                                                                                                         & Vukovic et al.~\cite{vukovic2009intelligent}~\textbf{(H)}                                                                                        & \begin{tabular}[c]{@{}c@{}}Jin et al.~\cite{jin2014news}~\textbf{(CA)}\\(hierarchical propagation model),\\ Chen et al.~\cite{chen2014email}~\textbf{(H)}\\(Levenshtein Distance)\end{tabular} \\ \bottomrule
\end{tabular}
}
\caption{Studies that focus on the detection of false information on OSNs. The table demonstrates the main methodology of each study, as well as the considered OSNs.  Also, we report the type of false information that is considered (see bold markers and cf. with Section~\ref{subsec:types}, \textbf{CA} corresponds to Credibility Assessment and refers to work that aim to assess the credibility of information).}
\label{tbl:detection_summary}
\end{table}

\subsubsection{Machine Learning}

\descr{Credibility Assessment.} 
Previous work leverage machine learning techniques to assess the credibility of information. 
Specifically, Castillo et al.~\cite{castillo2011information} analyze 2.5k trending topics from Twitter during 2010 to determine the credibility of information.
For labeling their data they utilize crowdsourcing tools, namely AMT, and propose the use of conventional machine learning techniques (SVM, Decision Trees, Decision Rules, and Bayes Networks) that take into account message-based, user-based, topic-based and propagation-based features. 
Gupta and Kumaraguru~\cite{gupta2012credibility} analyze tweets about fourteen high impact news events during 2011.
They propose the use of supervised machine learning techniques with a relevance feedback approach that aims to rank the tweets according to their credibility score.
AlRubaian et al.~\cite{alrubaian2015multistage} propose the use of a multi-stage credibility assessment platform that consists of a relative importance component, a classification component, and an opinion mining component.
The relative importance component requires human experts and its main objective is to rank the features according to their importance. 
The classification component is based on a Naive Bayes classifier, which is responsible for classifying tweets by taking the output of the relative importance component (ranked features), while the opinion mining component captures the sentiment of the users that interact with the tweets. 
The output of the three components is then combined to calculate an overall assessment.
Ma et al.~\cite{ma2015detect} observe that typically the features of messages in microblogs vary over time and propose the use of an SVM classifier that is able to consider the messages features in conjunction with how they vary over time. 
Their experimental evaluation, based on Twitter data provided by~\cite{castillo2011information} and on a  Sina Weibo dataset, indicate that the inclusion of the time-varying features increase the performance between 3\% and 10\%.

All of the aforementioned work propose the use of supervised machine learning techniques. 
In contrast, Gupta et al.~\cite{gupta2014tweetcred} propose a semi-supervised model that ranks tweets according to their credibility in real-time.
For training their model, they collect 10M tweets from six incidents during 2013, while they leverage CrowdFlower~\cite{crowdflower} to obtain groundtruth.
Their system also includes a browser extension that was used by approx. 1.1k users in a 3-month timespan, hence computing the credibility score of 5.4M tweets.
Their evaluation indicates that 99\% of the users were able to receive credibility scores under 6 seconds.
However, feedback from users for approx. 1.2k tweets indicate that 60\% of the users disagreed with the predicted score.

Volkova et al.~\cite{volkova2017separating} motivated by the performance gains of deep learning techniques, propose the use of neural networks to distinguish news into satire, hoaxes, clickbait, and propaganda news.
They collect 130k news posts from Twitter and propose the use of neural networks that use linguistic and network features.
Their findings indicate that Recurrent and Convolutional neural networks exhibit strong performance in distinguishing news in the aforementioned categories.

\descr{Rumors.} 
Kwon et al.~\cite{kwon2013prominent} propose the use of Decision Trees, Random Forest, and SVM for detecting rumors on Twitter.
Their models leverage temporal, linguistics, and structural features from tweets and can achieve precision and recall scores between 87\% and 92\%.
Yang et al.~\cite{yang2015emerging} propose the use of a hot topic detection mechanism that work in synergy with conventional machine learning techniques (Naive Bayes, Logistic Regression and Random Forest).
Liu et al.~\cite{liu2015real} demonstrate the feasibility of a real-time rumoring detection system on Twitter. 
To achieve real-time debunking of rumors, they propose the use of an SVM classifier that uses beliefs from the users in conjunction with traditional rumor features from~\cite{castillo2011information, yang2012automatic}.
Their evaluation demonstrates that for new rumors (5-400 tweets), the proposed classifier can outperform the models from~\cite{castillo2011information, yang2012automatic}. 
Furthermore, they compare their approach with human-based rumor debunking services (Snopes and Emergent), showing that they can debunk 75\% of the rumors earlier than the corresponding services.
Similarly, Kwon et al.~\cite{kwon2017rumor} study the rumor classification task with a particular focus on the temporal aspect of the problem, by studying the task over varying time windows on Twitter. 
By considering user, structural, linguistic, and temporal features, they highlight that depending on the time window, different characteristics are more important than others. 
For example, at early stages of the rumor propagation, temporal and structural are not available. 
To this end, they propose a rumor classification algorithm that achieves satisfactory accuracy both on short and long time windows.

Hamidian and Diab~\cite{hamidian2016rumor} propose a supevised model that is based on the Tweet Latent Vector (TLV), which is an 100-dimensional vector, proposed by the authors, that encapsulates the semantics behind a particular tweet. 
For the classification task, they use an SVM Tree Kernel model that achieves 97\% on two Twitter datasets. 
Giasemidis et al.~\cite{giasemidis2016determining} study 72 rumors in Twitter by identifying 80 features for classifying false and true rumors. 
These features include diffusion and temporal dynamics, linguistics, as well as user-related features. 
For classifying tweets, they use several machine learning techniques and conclude that Decision Trees achieve the best performance with an accuracy of 96\%.
Yang et al.~\cite{yang2012automatic} study the rumor detection problem in the Sina Weibo OSN. 
For the automatic classification task of the posts they use SVMs that take as input various features ranging from content-based to 
user- and location-based features.
Their evaluation shows that the classifier achieves an accuracy of approximately 78\%. 
Similarly to the aforementioned work, Wu et al.~\cite{wu2015false} try to tackle the rumor detection problem in the Sina Weibo OSN by leveraging SVMs.
Specifically, they propose an SVM classifier which is able to combine a
normal radial basis function, which captures high level semantic features, and a random walk graph kernel, which captures the similarities between propagation trees. 
These trees encompass various details such as temporal behavior, sentiment of re-posts, and user details.
Liang et al.~\cite{liang2015rumor} study the problem of rumor detection using machine learning solutions that take into account users' behavior in the Sina Weibo OSN. 
Specifically, they introduce 3 new features that are shown to provide up to 20\% improvement when compared with baselines. 
These features are: 
1) average number of followees per day;
2) average number of posts per day; 
and 3) number of possible microblog sources.
Zhang et al.~\cite{zhang2015automatic} propose various implicit features that can assist in the detection of rumors. 
Specifically, they evaluate an SVM classifier against the Sina Weibo dataset proposed in \cite{yang2012automatic} with the following features: 
1) content-based implicit features (sentiment polarity, opinion on comments and content popularity);
2) user-based implicit features (influence of user to network, opinion re-tweet influence, and match degree of messages) 
and 3) shallow message features that are proposed by the literature.
Their evaluation shows that the proposed sets of features can improve the precision and recall of the system by 7.1\% and 6.3\%, respectively.
Qin et al.~\cite{qin2016spotting} propose the use of a new set of features for detecting rumors that aim to increase the detection accuracy; namely novelty-based and pseudo-feedback features. 
The novelty-based features consider reliable news to find how similar is a particular rumor with reliable stories.
The pseudo-feedback features take into account information from historical confirmed rumors to find similarities. 
To evaluate their approach, they obtain messages from the Sina Weibo OSN and news articles from Xinhua News Agency~\cite{xinhuanet}. They compare an SVM classifier, which encompasses the aforementioned set of features and a set of other features (proposed by the literature), with the approaches proposed by~\cite{yang2012automatic, liu2015real}. 
Their findings indicate that their approach provides an improvement between 17\% and 20\% in terms of accuracy.
Similarly to~\cite{qin2016spotting}, Wu et al.~\cite{wu2017gleaning} propose a system that uses historical data about rumors for the detection task. 
Their system consists of a feature selection module, which categorizes and selects features, and a classifier.
For constructing their dataset they use Snopes and the Twitter API to retrieve relevant tweets, acquiring in total 10k tweets, which are manually verified by annotators.
In their evaluation, they compare their system with various baselines
finding that the proposed system offers enhanced performance in rumor detection with an increase of 12\%-24\% for precision, recall, and F1-score metrics.
Ma et al.~\cite{ma2016detecting} leverage Recurrent neural networks  to solve the problem of rumor detection in OSNs. 
Such techniques are able to learn hidden representations of the input without the need for hand-crafted features.
For evaluating their model, they construct two datasets; one from Twitter and one from Sina Weibo. 
For the labeling of their messages they use Snopes for Twitter and the official rumor-busting service of Sina Weibo's OSN.
Their evaluation shows an accuracy of 91\% on the Sina Weibo dataset and 88\% on the Twitter dataset.

\descr{Hoaxes.} Tacchini et al.~\cite{tacchini2017some} study hoaxes in Facebook and argue that they can accurately discern hoax from non-hoax posts by simply looking at the users that liked the posts. 
Specifically, they propose the use of Logistic Regression that classifies posts with features based on users' interactions. 
Their evaluation demonstrate that they can identify hoaxes with an accuracy of 99\%.
Kumar et al.~\cite{kumar2016disinformation} study the presence of hoaxes in Wikipedia articles by considering 20k hoax articles that are explicitly flagged by Wikipedia editors. 
They find that most hoaxes are detected quickly and have little impact, however, a small portion of these hoaxes have a significant life-span and are referenced a lot across the Web. 
By comparing the "successful" hoaxes with failed hoaxes and legitimate articles, the authors highlight that the successful hoaxes have notable differences in terms of structure and content.
To this end, they propose the use of a Random Forest classifier to distinguish if articles are hoaxes. 
Their evaluation reports that their approach achieves an accuracy of 92\% and that is able to outperform human judgments by a significant margin (20\%). 
Maigrot et al.~\cite{maigrot2016mediaeval} propose the use of a multi-modal hoax detection system that fuses the diverse modalities pertaining to a hoax.
Specifically, they take into consideration the text, the source, and the image of tweets.
They observe higher performance when using only the source or text modality instead of the combination of all modalities.

\descr{Conspiracy Theories.} Conti et al.~\cite{conti2017s} focus on identifying conspiracy theories in OSNs by considering only the structural features of the information cascade.
The rationale is that such features are difficult to be tampered by malicious users, which aim to avoid detection from classification systems.   
For their dataset they use data from~\cite{bessi2015science}, which consist of scientific articles and conspiracy theories. 
For classifying their Facebook data they propose conventional machine learning techniques and they find that it is hard to distinguish a conspiracy theory from a scientific article by only looking at their structural dynamics (F1 -score not exceeding 65\%). 

\descr{Satire.} Rubin et al.~\cite{rubin2016fake} propose the use of satirical cues for the detection of false information on news articles.
Specifically, they propose the use of five new set of features, namely absurdity, humor, grammar, negative affect, and punctuation.
Their evaluation shows that by using an SVM algorithm with the aforementioned set of features and others proposed by the literature, they can detect satirical news with 90\% precision and 84\% recall. 

\descr{Clickbait.} Several studies focus on the detection of clickbait on the Web using machine learning techniques. Specifically, Chen et al.~\cite{chen2015misleading} propose tackling the problem using SVMs and Naive Bayes. 
Also, Chakraborty et al.~\cite{chakraborty2016stop} propose the 
use of SVM and a browser add-on to offer a system to users for news articles. 
Potthast et al.~\cite{potthast2016clickbait} proposes the use of Random Forest for detecting clickbait tweets.
Moreover, Biyani et al.~\cite{Biyani2016} propose the use of Gradient Boosted Decision Trees for clickbait detection in news articles and show that the degree of informality in the content of the landing page can help in finding clickbait articles. 
Anand et al.~\cite{anand2016we} is the first work that suggests the use of deep learning techniques for mitigating the clickbait problem.
Specifically, they propose the use of Recurrent Neural Networks in conjunction with word2vec embeddings~\cite{mikolov2013distributed} for identifying clickbait news articles.
 Similarly, Zannettou et al.~\cite{zannettou2018good} use deep learning techniques to detect clickbaits on YouTube. Specifically, they  propose a semi-supervised model based on variational autoencoders (deep learning). Their evaluation indicates that they can detect clickbaits with satisfactory performance and that YouTube's recommendation engine does not consider clickbait videos in its recommendations. 

\descr{Fabricated.} Wang~\cite{wang2017liar} presents a dataset that consists of 12.8k manually annotated short statements obtained from PolitiFact.
They propose the use of Convolutional neural networks for fusing linguistic features with metadata (e.g., who is the author of the statement). 
Their evaluation demonstrates that the proposed model outperforms SVM and Logistic Regression algorithms.

\subsubsection{Systems}
\descr{Rumors.} Resnick et al.~\cite{resnick2014rumorlens} propose a system called RumorLens, which aims to discover rumors in a timely manner, provide insights regarding the rumor's validity, and visualize a rumor's propagation. 
To achieve the aforementioned, RumorLens leverages data mining techniques alongside with a visual analysis tool.
However, their system raises scalability issues as it highly depends on users' labor, which provide labeling of tweets that are subsequently used for classifying tweets related to a particular rumor.
Vosoughi et al.~\cite{vosoughi2015human} propose a human-machine collaborative system that aims to identify rumors by disposing irrelevant data and ranking the relevant data.
Their system consists of two components; the assertion detector and the hierarchical clustering module. 
The assertion detector is a classifier that uses semantic and syntactic features to find tweets that contain assertions. 
These tweets are then presented to the clustering module, which clusters the tweets according to the similarity of the assertions.
During their evaluation, the authors state that for a particular incident (Boston Marathon Bombings) from a dataset of 20M tweets, their system managed to discard 50\% of them using the assertion detector. 
Furthermore, the 10M relevant tweets are clustered somewhere between 100 and 1000 clusters, something that enables users to quickly search and find useful information easier.

\descr{Credibility Assessment.} Jaho et al.~\cite{jaho2014alethiometer} undertake a statistical analysis by crawling Twitter for 3 months and retrieve a dataset that includes 10M users.
They propose a system that is based on contributor-related features (e.g., reputation, influence of source, etc.), content features (e.g., popularity, authenticity, etc.) and context features (e.g., coherence, cross-checking, etc.). 
Their system combines all the features and outputs a single metric that corresponds to the truthfulness of the message.
Zhou et al.~\cite{zhou2015real} note that calculating credibility in the granularity of message is not scalable, therefore they propose the calculation of credibility score per event. 
To this end, they propose a system that is able to collect related data from Sina Weibo using keywords and detect the credibility of a particular event.
The credibility score is calculated by the combination of 3 sub-models; the user model, the propagation model, and the content model. 
Each one of the sub-models considers one aspect of the news and the overall score is calculated using weighted combination.
The system is trained on a dataset that contains 73 real news and 73 fake news from approximately 50k posts.
Their evaluation shows that the proposed system provides an accuracy close to 80\% and that credibility scores are calculated within 35 seconds.

\descr{Hoaxes.} Vukovic et al.~\cite{vukovic2009intelligent} focus on hoaxes and  propose the use of a detection system for email.
The proposed system consists of a feed-forward neural network and a self-organizing map (SOM) and it is trained on a corpus of 298 hoax and 1370 regular emails.
The system achieves an accuracy of 73\% with a ratio of false positives equal to 4.9\%.
Afroz et al.~\cite{afroz2012detecting} focus on detecting hoaxes by observing changes in writing style.
The intuition is that people use different linguistic features when they try to obfuscate or change information from users.
To detect hoaxes they propose the use of an SVM classifier that takes into account the following set of features: 
1) lexical features;
2) syntactic features;
3) content features 
and 4) lying detection features obtained from~\cite{burgoon2003detecting, hancock2007lying}.
Their evaluation on various datasets indicates that the proposed system can detect hoaxes with an accuracy of 96\%.

\subsubsection{Other models/algorithms}

\descr{Rumors.} Qazvinian et al.~\cite{qazvinian2011rumor} study the rumor detection problem on Twitter by retrieving tweets regarding rumors and leveraging manual inspectors to annotate it. 
Specifically, the annotators were asked whether tweets contained rumors or not and whether a user endorsed, debunked or was neutral about the rumors. 
The resulted dataset consists of approximately 10k annotated tweets and was analyzed to demonstrate the effectiveness of the following feature sets in identifying rumors: 1) content-based features; 2) network-based features and 3) Twitter-specific memes (hashtags and URLs). 
Furthermore, the paper proposes a rumor retrieval model that achieves 95\% precision. %
Zhao et al.~\cite{zhao2015enquiring} are motivated by the fact that identifying false factual claims in each individual message is intractable. 
To overcome this, they adapt the problem in finding whole clusters of messages that their topic is a disputed factual claim. 
To do so, they search within posts to find specific phrases that are used from users who want to seek more information or to express their skepticism. For example, some enquiry phrases are "Is this true?", "Really?" and "What?". 
Their approach uses statistical features of the clusters in order to rank them according to the likelihood of including a disputed claim.
Their evaluations on real Twitter data indicate that among the top 50 ranked clusters, 30\% of them are confirmed rumors.

\descr{Fabricated.} Farajtabar et al.~\cite{farajtabar2017fake} propose a framework for tackling false information that combines a multivariate Hawkes process and reinforcement learning. 
Their evaluation highlights that their model shows promising performance in identifying false information in real-time on Twitter.
Kumar and Geethakumari~\cite{kumar2014detecting} measure the diffusion of false information by exploiting cues obtained from cognitive psychology. 
Specifically, they consider the consistency of the message, the coherency of the message, the credibility of the source, and the general acceptability of the content of the message. 
These cues are fused to an algorithm that aims to detect the spread of false information as soon as possible.
Their analysis on Twitter reports that the proposed algorithm has a 90\% True positive rate and a False positive rate less than 10\%.

\descr{Credibility Assessment.} Jin et al.~\cite{jin2016news} aim to provide verification of news by considering conflicting viewpoints on Twitter and Sina Weibo. 
To achieve this, they propose the use of a topic model method that identifies conflicting viewpoints.
Subsequently they construct a credibility network with all the viewpoints and they formulate the problem as a graph optimization problem, which can be solved with an iterative approach.
They compare their approach with baselines proposed in~\cite{castillo2011information, kwon2013prominent}, showing that their solution performs better.
Jin et al.~\cite{jin2014news} propose a hierarchical propagation model to evaluate information credibility in microblogs by detecting events, sub-events, and messages.
This three-layer network assists in revealing vital information regarding information credibility. 
By forming the problem as a graph optimization problem, they propose an iterative algorithm, that boosts the accuracy by 6\% when compared to an SVM classifier that takes into account only features obtained from the event-level network only.

\descr{Biased.} Potthast et al.~\cite{potthast2017stylometric} study the writing style of hyperpartisan news (left-wing and right-wing) and mainstream news and how this style can be applied in hyperpartisan news detection.
Their dataset consists of 1.6k news articles from three right-wing, three left-wings, and three mainstream news sites. 
For annotating the dataset they used journalists from Buzzfeed, who rated each article according to its truthfulness.
By leveraging the Unmasking approach \cite{koppel2007measuring}, the paper demonstrates that right-wing and left-wing hyperpartisan news exhibit similar writing style that differentiates from the mainstream news.
To this end, they propose the use of Random Forest classifier that aims to distinguish hyperpartisanship.
Their evaluation indicates that their style-based classifier can distinguish hyperpartisan news with an accuracy of 75\%. 
However, when the same classifier is used to discern fake or real news, then the accuracy is 55\%. 

\descr{Hoaxes.} Chen et al.~\cite{chen2014email} propose an email hoax detection system by incorporating a text matching method using the Levenshtein distance measure. 
Specifically, their system maintains a database of hoaxes that is used to calculate the distance between a potential hoax email and the stored hoaxes.

\subsection{Containment of false information}

Several studies focus on containing the diffusion of false information. 
Our literature review reveals that the majority of previous work on containment of rumors, while we also find one that focus on Hoaxes (see Tambuscio et al.~\cite{tambuscio2015fact}). 
Below we provide a brief overview of the studies that try to contain the spread of false information, while ensuring that the solutions are scalable.

\descr{Rumors.}
Tripathy et al. ~\cite{tripathy2010study} propose a process, called "anti-rumor", which aims to mitigate the spreading of a rumor in a network. 
This process involves the dissemination of messages, which contradict with a rumor, from agents. 
The authors make the assumption that once a user receives an anti-rumor message, then he will never believe again the rumor, thus the spreading of a rumor is mitigated. 
Their evaluation, based on simulations, indicates the efficacy of the proposed approach.
Budak et al.~\cite{budak2011limiting} formulate the problem of false information spreading as an optimization problem.
Their aim is to identify a subset of the users that need to be convinced to spread legitimate messages in contrast with the bad ones that spread rumors.
The paper shows that this problem is NP-hard and they propose a greedy solution as well as some heuristics to cope with scalability issues. 
Fan et al.~\cite{fan2013least} try to tackle the problem of false information propagation under the assumption that rumors originate from a particular community in the network. 
Similarly to other work, the paper tries to find a minimum set of individuals, which are neighbors with the rumor community to stop the rumor diffusion in the rest of the network. 
To achieve this, they propose the use of two greedy-based algorithms, which are evaluated in two real-world networks (Arxiv Hep and Enron). 
Their experimental results show that the proposed algorithms outperform simple heuristics in terms of the number of infected nodes in the network. 
However, as noted, the greedy algorithms are time consuming and are not applicable in large-scale networks.
Kotnis et al.~\cite{kotnis2014cost} propose a solution for stopping the spread of false information by training a set of individuals in a network that aim to distinguish and stop the propagation of rumors. 
This set of individuals is selected based on their degree in the network with the goal to minimize the overarching training costs. 
For evaluating their solution they create a synthetic network, which takes into account a calculated network degree distribution, based on ~\cite{molloy1995critical}.
Ping et al.~\cite{ping2014sybil} leverage Twitter data to demonstrate that sybils presence in OSNs can decrease the effectiveness of community-based rumor blocking approaches by 30\%. 
To this end, they propose a Sybil-aware rumor blocking approach, which finds a subset of nodes to block by considering the network structure in conjunction with the probabilities of nodes being sybils. 
Their evaluation, via simulations on Twitter data, show that the proposed approach significantly decreases the number of affected nodes, when compared to existing approaches.
He et al.~\cite{he2015modeling} argue that existing false information containment approaches have different costs and efficiencies in different OSNs.
To this end, they propose an optimization method that combines the spreading of anti-rumors and the block of rumors from influential users.
The goal of their approach is to minimize the overarching cost of the method while containing the rumor within an expected deadline.
To achieve this, they use the Pontryagin's maximum principle~\cite{kopp1962pontryagin} on the Digg2009 dataset \cite{hogg2012social}. 
They find that spreading the truth plays a significant role at the start of the rumor propagation, whereas close to the deadline of containment the blocking of rumors approach should be used extensively.
Huang et al.~\cite{huang2015connected} aim to contain the false information spread by finding and decontaminating with good information,
the smallest set of influential users in a network. 
To do so, they propose a greedy algorithm and a community-based heuristic, which takes into consideration the community structure of the underlying network. 
For evaluating their approach, they used traces from three networks; NetHEPT, NetHEPT\_WC and Facebook.
Previous studies on false information containment~\cite{budak2011limiting, nguyen2012containment} assumed that when true and false information arrive the same time at a particular node, then the true information dominates.
Wang et al.~\cite{wang2014containment} state that the dominance of the information should be based on the influence of the neighbors in the network. 
With this problem formulation in mind, the paper proposes two approaches to find the smallest number of nodes that are required to stop the false information spread. 
Their evaluation is based on three networks obtained from Twitter, Friendster, and a random synthetic network. 
Evaluation comparisons with simple heuristics (random and high degree) demonstrate the performance benefits of the proposed approaches.
In a similar notion, Tong et al.~\cite{tong2017efficient} aim to increase performance motivated by the fact that greedy solutions, which include Monte Carlo simulations, are inefficient as they are computationally intensive. 
To overcome this, the paper proposes a random-based approach, which utilizes sampling with the aim to be both effective and efficient. 
The performance evaluations on real-world (obtained from Wikipedia and Epinions~\cite{epinions}) and synthetic networks demonstrate that the proposed solution can provide a 10x speed-up without compromising performance when compared to state-of-the-art approaches.  
Wang et al.~\cite{wang2016drimux} propose a model, called DRIMUX, which aims to minimize the influence of rumors by blocking a subset of nodes while considering users' experience.
User experience is defined as a time threshold that a particular node is willing to wait while being blocked.
Their model utilizes survival theory and takes into account global rumor popularity features, individual tendencies (how likely is a rumor to propagate between a pair of nodes) as well as the users' experience.
Their evaluations on a Sina Weibo network, which consists of 23k nodes and 183k edges, indicate that the proposed model can reduce the overarching influence of false information.

\descr{Hoaxes.} Tambuscio et al.~\cite{tambuscio2015fact} simulate the spread and debunking of hoaxes on networks. 
Specifically, they model the problem as a competition between believers (acknowledge the hoax) and fact checkers which reveal the hoax with a specific probability. 
To study their model they performed simulations on scale-free and random networks finding that a specific threshold for the probability of fact checkers exists and this indicates that the spread can be stopped with a specific number of fact checkers. 
However, the paper oversimplifies the problem by assuming all the nodes to have the same probability.

\subsection{Detection and Containment of False Information - Future Directions}

The main findings from the literature review of the detection and containment of false information are:
1) Machine learning techniques can assist in identifying false information. However, they heavily rely on handcrafted set of features and it is unclear if they generalize well on other datasets;
2) Containment of false information can be achieved by adding a set of good nodes that disseminate good information or information that refute false; and
3) The problem of detection of false information requires human-machine collaboration for effectively mitigating it.

Despite the fact that several studies exist attempting to detect and contain false information on the Web, the problem is still emerging and prevalent.
This is mainly because the problem requires higher cognitive and context awareness that current systems do not have.
To assist in achieving a better detection of false information on the Web, we foresee the following tangible research directions.

First, information on the Web exists in multiple formats, and thus false information is disseminated via textual claims, screenshots, videos, etc.
Most studies, however, take into consideration only one format.
To achieve a multi-format false information detection system requires correlating information in multiple formats, which in turn requires understanding the similarities and differences in the content each format delivers.
To the best of our knowledge, a system that can meaningfully assist in detecting false information across multiple formats does not exist.
Next, to the best of our knowledge, no prior work has rigorously assessed credibility based on user profiling.
For example, a post from an expert on a particular subject should not be treated with the same weight as a post by a typical user.
We foresee studying false information detection from a users' perspective is a way forward in effectively detecting and containing the spread of false information on the Web.
Finally, most previous studies focus on detection and containment of false information on a single OSN, but the Web is much larger than any single platform or community.
Therefore, future work should address the problem with a holistic view of the information ecosystem.
This requires an understanding of how information jumps from one Web community to another, how Web communities influence each other, and how to correlate accounts that exist in multiple Web communities (e.g., how to find that a particular Facebook and Twitter account belong to the same user).
Such an understanding will be particularly useful for containing the spread of false information from one Web community to another.

\section{False Information in the political stage}
\label{sec:political}

Recently, after the 2016 US elections, the problem of false information dissemination got extensive interest from the community. Specifically, Facebook got openly accused for disseminating false information and that affected the outcome of the elections~\cite{facebook_fake_news_us_elections}. 
It is evident that dissemination of false information on the Web is used a lot for political influence.
Therefore in this section we review the most relevant studies on the political stage. Table~\ref{tbl:political_overview} reports the reviewed work as well as the main methodology and considered OSN.

\subsection{Machine Learning}

\descr{Propaganda.} Ratkiewicz et al.~\cite{ratkiewicz2011detecting} study political campaigns on Twitter that use multiple controlled accounts to disseminate support for an individual or opinion. 
They propose the use of a machine learning-based framework in order to detect the early stages of the spreading of political false information on Twitter.
Specifically, they propose a framework that takes into consideration topological, content-based and crowdsourced features of the information diffusion in Twitter.
Their experimental evaluation demonstrates that the proposed framework achieves more than 96\% accuracy in the detection of political campaigns for data pertaining to the 2010 US midterm elections.
Conover et al.~\cite{conover2011political} study Twitter on a six-week period leading to the 2010 US midterm elections and the interactions between right and left leaning communities. 
They leverage clustering algorithms and manually annotated data to create the re-tweets and mentions networks.
Their findings indicate that the re-tweet network has limited connectivity between the right and left leaning communities, whereas this is not the case in the mentions networks.
This is because, users try to inject different opinions on users with different ideologies, by using mentions on tweets, so that they change their stance towards a political individual or situation.
Ferrara et al.~\cite{ferrara2016detection} propose the use of a k-nearest neighbor algorithm with a dynamic warping classifier in order to capture promoted campaigns in Twitter. 
By extracting a variety of features (user-related, timing-related, content-related and sentiment-related features) from a large corpus of tweets they demonstrate that they can distinguish promoted campaigns with an AUC score close to 95\% in a timely manner.
 
 \begin{table}[]
\centering
\resizebox{0.8\columnwidth}{!}{
\begin{tabular}{@{}cccc@{}}
\toprule
\textbf{Platform} & \textbf{Machine Learning}                                                                                                                                                               & \textbf{OSN Data Analysis}                                                                                                                                                                                                                                                                                                                                                                   & \textbf{Other models/algorithms}                                                                                                                                                                                                                                                                                                                                                                                        \\ \midrule
Twitter           & \begin{tabular}[c]{@{}c@{}} Ratkiewicz et al.~\cite{ratkiewicz2011detecting}~\textbf{(P)},\\ Conover et al.\cite{conover2011political}~\textbf{(P)},\\ Ferrara et al.\cite{ferrara2016detection}~\textbf{(P)}\end{tabular} & \begin{tabular}[c]{@{}c@{}} Wong et al.~\cite{wong2013quantifying}~\textbf{(B)},\\ Golbeck and Hansen~\cite{golbeck2014twitterpolpref}~\textbf{(B)},\\ Jackson and Welles~\cite{jackson2015hijacking}~\textbf{(P)},\\ Hegelich and Janetzko\cite{hegelich2016social}~\textbf{(P)},\\Zannettou et al.~\cite{zannettou2018disinformation}~\textbf{(P)}\\ Howard and Kollanyi\cite{howard2016botsa}~\textbf{(P)},\\ Shin et al.\cite{shin2016political}~\textbf{(R)}\end{tabular} & \begin{tabular}[c]{@{}c@{}}An et al.~\cite{an2012visualizing}~\textbf{(B)} \\(distance model),\\ Al-khateeb and Agarwal~\cite{al2015examining}~\textbf{(P)} \\(social studies)\\ Ranganath et al.\cite{ranganath2016understanding}~\textbf{(P)} \\(exhaustive search),\\ Jin et al.~\cite{jin2017rumor}~\textbf{(R)} \\(text similarity),\\ Yang et al.~\cite{yang2016social}~\textbf{(B)} \\(agenda-setting tool)\end{tabular} \\ \midrule
Digg              & Zhou et al.\cite{zhou2011classifying}~\textbf{(B)}                                                                                                                                                 & X                                                                                                                                                                                                                                                                                                                                                                                            & X                                                                                                                                                                                                                                                                                                                                                                                                                       \\ \midrule
Sina Weibo        & X                                                                                                                                                                                       & \begin{tabular}[c]{@{}c@{}}King et al.~\cite{king2016chinese}~\textbf{(P)},\\ Yang et al.~\cite{yang2015penny}~\textbf{(P)}\end{tabular}                                                                                                                                                                                                                                                                             & X                                                                                                                                                                                                                                                                                                                                                                                                                       \\ \midrule
News articles     & Budak et al.~\cite{budak2016fair}~\textbf{(B)}                                                                                                                                                      & Woolley\cite{woolley2016automating}~\textbf{(P)}                                                                                                                                                                                                                                                                                                                                                        & X                                                                                                                                                                                                                                                                                                                                                                                                                       \\ \midrule
Facebook          & X                                                                                                                                                                                       & Allcot and Gentzkow\cite{allcott2017social}~\textbf{(P)}                                                                                                                                                                                                                                                                                                                                                & X                                                                                                                                                                                                                                                                                                                                                                                                                       \\ \bottomrule
\end{tabular}
}
\caption{Studies on the false information ecosystem on the political stage. The table demonstrates the main methodology of each study as well as the considered OSNs.}
\label{tbl:political_overview}
\end{table}

\descr{Biased.}
Zhou et al.~\cite{zhou2011classifying} study Digg, a news aggregator site, and aim to classify users and articles to either liberal or conservative.
To achieve this, they propose three semi-supervised propagation algorithms that classify users and articles based on users' votes.
The algorithms make use of a few labeled users and articles to predict a large corpus of unlabeled users and articles.
The algorithms are based on the assumption that a liberal user is more likely to vote for a liberal article rather than a conservative article. 
Their evaluations demonstrate that the best algorithm achieves 99\% and 96\% accuracy on the dataset of users and articles, respectively.
Budak et al.~\cite{budak2016fair} use Logistic Regression to identify articles regarding politics from a large corpus of 803K articles obtained from 15 major US news outlets.
Their algorithm filtered out 86\% of the articles as non-political related, while a small subset of the remainder (approx. 11\%) were presented to workers on AMT. 
The workers were asked to answer questions regarding the topic of the article, whether the article was descriptive or opinionated, the level of partisanship, and the level of bias towards democrats or republicans.
Their empirical findings are that on these articles there are no clear indications of partisanship, some articles within the same outlet are left-leaning and some have right-leaning, hence reducing the overall outlet bias. 
Also, they note that usually bias in news articles is expressed by criticizing the opposed party rather than promoting the supporting party.

\subsection{OSN Data Analysis}
\descr{Biased.}
Wong et al.~\cite{wong2013quantifying} collect and analyze 119M tweets pertaining to the 2012 US presidential election to quantify political leaning of users and news outlets. 
By formulating the problem as an ill-posed linear inverse problem, they propose an inference engine that considers tweeting behavior of articles.
Having demonstrated their inference engine, the authors report results for the political leaning scores of news sources and users on Twitter.
Golbeck and Hansen~\cite{golbeck2014twitterpolpref} provide a technique to estimate audience preferences in a given domain on Twitter, with a particular focus on political preferences.
Different from methods that assess audience preference based on citation networks of news sources as a proxy, they directly measure the audience itself via their social network.
Their technique is composed of three steps: 1)~apply ground truth scores (they used Americans for Democratic Action reports as well as DW-Nominate scores) to a set of seed nodes in the network, 2)~map these scores to the seed group's followers to create ``P-scores'', and 3)~map the P-scores to the target of interest (e.g., government agencies or think tanks).
One important take away from this work is that \emph{Republicans are over-represented on Twitter with respect to their representation in Congress}, at least during the 2012 election cycle.
To deal with this, they built a balanced dataset by randomly sampling from bins formed by the number of followers a seed group account had.

\descr{Propaganda.} Jackson and Welles~\cite{jackson2015hijacking} demonstrate how Twitter can be exploited to organize and promote counter narratives.
To do so, they investigate the misuse of a Twitter hashtag (\#myNYPD) during the 2014 New York City Police Department public relations campaign.
In this campaign, this hashtag was greatly disseminated to promote counter narratives about racism and police misconduct.
The authors leverage network and qualitative discourse analysis to study the structure and strategies used for promoting counterpublic narratives.

Hegelich and Janetzko~\cite{hegelich2016social} investigate whether bots on Twitter are used as political actors.
By exposing and analyzing 1.7K bots on Twitter, during the Russian/Ukrainian conflict, they find that the botnet has a political agenda and that bots exhibit various behaviors.
Specifically, they find that bots try to hide their identity, to be interesting by promoting topics through the use of hashtags and retweets.
Howard and Kollanyi~\cite{howard2016botsa} focus on the 2016 UK referendum and the role of bots in the conversations on Twitter.
They analyze 1.5M tweets from 313K Twitter accounts collected by searching specific hashtags related to the referendum.
Their analysis indicates that most of the tweets are in favor of exiting the EU, there are bots with different levels of automation and that 1\% of the accounts generate 33\% of the overall messages.
They also note that among the top sharers, there are a lot of bot accounts that are mostly retweeting and not generating new content. 
In a similar work, Howard et al.~\cite{howard2016botsb} study Twitter behavior during the second 2016 US Presidential Debate. They find that Twitter activity is more pro-Trump and that a lot of activity is driven by bots.
However, they note that a substantial amount of tweets is original content posted from regular Twitter users.
Woolley~\cite{woolley2016automating} analyzes several articles regarding the use of bots in OSNs for political purposes.
Specifically, he undertakes a qualitative content analysis on 41 articles regarding political bots from various countries obtained from the Web. %
One of his main findings is that the use of bots varies from country to country and that some countries (e.g., Argentina, China, Russia, USA, etc.) use political bots on more than one type of event.
For example, they report the use of Chinese political bots for elections, for protests and for security reasons.

In the Chinese political stage, during December 2014, an anonymous blogger released an archive of emails pertaining to the employment of Wumao, a group of people that gets paid to disseminate propaganda on social media, from the Chinese government.
King et al.~\cite{king2016chinese} analyzed these leaks and found out 43K posts that were posted by Wumao. 
Their main findings are:
1) by analyzing the time-series of these posts, they find bursty activity, hence signs of coordination of the posters;
2) most of the posters are individuals working for the government; and 
3) by analyzing the content of the message, they note that posters usually post messages for distraction rather than discussions of controversial matters (i.e., supporting China's regime instead of discussing an event).
Similarly to the previous work, Yang et al.~\cite{yang2015penny} study the Wumao by analyzing 26M posts from 2.7M users on the Sina Weibo OSN, aiming to provide insights regarding the behavior and the size of Wumao.
Due to the lack of ground truth data, they use clustering and topic modeling techniques, in order to cluster users that post politics-related messages with similar topics.
By manually checking the users on the produced clusters, they conclude that users that post pro-government messages are distributed across multiple clusters, hence there is no signs of coordination of the Wumao on Sina Weibo for the period of their dataset (August 2012 and August 2013).

Zannettou et al.~\cite{zannettou2018disinformation} study Russian state-sponsored troll accounts and measure the influence they had on Twitter and other Web communities. They find that Russian trolls were involved in the discussion of political events, and that they exhibit different behavior when compared to random users.
Finally, they show that their influence was not substantial, with the exception of the dissemination of articles from state-sponsored Russian news outlets like Russia Today (RT).
Allcot and Gentzkow~\cite{allcott2017social} make a large scale analysis on Facebook during the period of the 2016 US election.
Their results provide the following interesting statistics about the US election:
1) 115 pro-Trump fake stories are shared 30M times, whereas 41 pro-Clinton fake stories are shared 7.6M times. This indicates that fake news stories that favor Trump are more profound in Facebook.
2) The aforementioned 37.6M shares translates to 760M instances of a user clicking to the news articles. This indicates the high reachability of the fake news stories to end-users. 
3) By undertaking a 1200-person survey, they highlight that a user's education, age and overall media consumption are the most important factors that determine whether a user can distinguish false headlines.

\descr{Rumors.} Shin et al.~\cite{shin2016political} undertake a content-based analysis on 330K tweets pertaining to the 2012 US election.
Their findings agree with existing literature, noting that users that spread rumors are mostly sharing messages against a political person.
Furthermore, they highlight the resilience of rumors despite the fact that rumor debunking evidence was disseminated in Twitter;  however, this does not apply for rumors that originate from satire websites.

\subsection{Other models/algorithms}

\descr{Biased.}
An et al.~\cite{an2012visualizing} study the interactions of 7M followers of 24 US news outlets on Twitter, in order to identify political leaning.
To achieve this, they create a distance model, based on co-subscription relationships, that maps news sources to a dimensional dichotomous political spectrum.
Also, they propose a real-time application, which utilizes the underlying model, and visualizes the ideology of the various news sources.
Yang et al.~\cite{yang2016social} investigate the topics of discussions on Twitter for 51 US political persons, including President Obama.
The main finding of this work is that Republicans and Democrats are similarly active on Twitter with the difference that Democrats tend to use hashtags more frequently.
Furthermore, by utilizing a graph that demonstrates the similarity of the agenda of each political person, they highlight that Republicans are more clustered.
This indicates that Republicans tend to share more tweets regarding their party's issues and agenda.

\descr{Propaganda.}
Al-khateeb and Agarwal~\cite{al2015examining} study the dissemination of propaganda on Twitter from terrorist organizations ( namely ISIS).
They propose a framework based on social studies that aim to identify social and behavioral patterns of propaganda messages disseminated by a botnet. 
Their main findings are that bots exhibit similar behaviors (i.e., similar sharing patterns, similar usernames, lot of tweets in a short period of time) and that they share information that contains URLs to other sites and blogs.
Ranganath et al.~\cite{ranganath2016understanding} focus on the detection of political advocates (individuals that use social media to strategically push a political agenda) on Twitter. 
The authors note that identifying advocates is not a straightforward task due to the nuanced and diverse message construction and propagation strategies.
To overcome this, they propose a framework that aims to model all the different propagation and message construction strategies of advocates.
Their evaluation on two datasets on Twitter regarding gun rights and elections demonstrate that the proposed framework achieves good performance with a 93\% AUC score.

\descr{Rumors.} Jin et al.~\cite{jin2017rumor} study the 2016 US Election through the Twitter activity of the followers of the two presidential candidates.
For identifying rumors, they collect rumor articles from Snopes and then they use text similarity algorithms based on:
1) Term frequency-inverse document frequency (TF-IDF);
2) BM25 proposed in~\cite{robertson2009probabilistic}
3) Word2Vec embeddings~\cite{mikolov2013distributed};
4) Doc2Vec embeddings~\cite{le2014distributed};
5) Lexicon used in ~\cite{zhao2015enquiring}.
Their evaluation indicates that the best performance is achieved using the BM25-based approach.
This algorithm is subsequently used to classify the tweets of the candidates' followers.
Based on the predictions of the algorithm, their main findings are:
1) rumors are more prevalent during election period;
2) most of the rumors are posted by a small group of users;
3) rumors are mainly posted to debunk rumors that are against their presidential candidate, or to inflict damage on the other candidate; and
4) rumor sharing behavior increases in key points of the presidential campaign and in emergency events.

\subsection{False information in political stage - Future Directions}
The main insights from the review of work that focus on the political stage are:
1) Temporal analysis can by leveraged to assess coordination of bots, state-sponsored actors, and orchestrated efforts on disseminating political false information;
2) Bots are extensively used for the dissemination of political false information;
3) Machine learning techniques can assist in detecting political false information and political leaning of users. However, there are concerns about the generalization of such solutions on other datasets/domains; and
4) Political campaigns are responsible for the substantial dissemination of political false information in mainstream Web communities.

As future directions for understanding and mitigating the effects of false information on the Web, we propose the following.
First, there is extensive anecdotal evidence highlighting that Web communities are used by state-sponsored troll factories, e.g., the recent news regarding Russian troll factories deployed to influence the outcomes of the 2016 Brexit~\cite{brexit_russians} referendum and the 2016 US presidential election~\cite{us_elections_russians}.
We thus propose investigating this phenomenon both from the perspective of user analytics as well as societal impact.
Additionally, there is a lack of studies providing insight on how politics-related false information is disseminated across the Web; most studies focus on a single Web community or to specific events or do not examine politics.
Understanding how political information propagates across the Web will help society identify the source of false information and lead to successful containment efforts.

\section{Other related work}
\label{sec:other}

In this section we shall present work that is relevant to the false information ecosystem that does not fit in the aforementioned lines of work.
Specifically, we group these studies in the following categories:
1) General Studies;
2) Systems; and
3) Use of images on the false information ecosystem.

\subsection{General Studies}

\descr{Credibility Assessment.} Buntain and Golbeck~\cite{buntain2017want} compare the accuracy of models that use features based on journalists assessments and crowdsourced assessments.
They indicate that there is small overlap between the two features sets despite the fact that they provide statistically correlated results. 
This indicates that crowdsourcing workers discern different aspects of the stories when compared to journalists.
Finally, they demonstrate that models that utilize features from crowdsourcing outperform the models that utilize features from journalists.
Zhang et al.~\cite{zhang2018structured} present a set of indicators that can used to assess the credibility of articles. 
To find these indicators they use a diverse set of experts (coming from multiple disciplines), which analyzed and annotated 40 news articles.
Despite the low number of annotated articles, this inter-disciplinary study is important as it can help in defining standards for assessing the credibility of content on the Web.
Mangolin et al.~\cite{margolin2018political} study the interplay between fact-checkers and rumor spreaders on social networks finding that users are more likely to correct themselves if the correction comes from a user they follow when compared to a stranger.

\descr{Conspiracy Theories.} Starbird~\cite{starbird2017examining} performs a qualitative analysis on Twitter regarding shooting events and conspiracy theories.
Using graph analysis on the domains linked from the tweets, she provides insight on how various websites work to promote conspiracy theories and push political agendas.

\descr{Fabricated.} Horne and Adah~\cite{horne2017just} focus on the headline of fake and real news. 
Their analysis on three datasets of news articles highlight that fake news have substantial differences in their structure when compared with real news. 
Specifically, they report that generally the structure of the content and the headline is different.
That is, fake news are smaller in size, use simple words, and use longer and ``clickbaity'' headlines. 
Potts et al.~\cite{potts2013interfaces} study Reddit and 4chan and how their interface is a part of their culture that affects their information sharing behavior.
They analyze the information shared on these two platforms during the 2013 Boston Marathon bombings. 
Their findings highlight that users on both sites tried to find the perpetrator of the attack by creating conversations for the attack, usually containing false information.
Bode and Vraga~\cite{bode2015related} propose a new function on Facebook, which allow users to observe related stories that either confirm or correct false information; they highlight that using this function users acquire a better understanding of the information and its credibility.
Finally, Pennycook and Rand~\cite{pennycook2017implied} highlight that by attaching warnings to news articles can help users to better assess the credibility of articles, however news articles that are not attached with warnings are considered as validated, which is not always true, hence users are tricked.

\descr{Propaganda.} Chen et al.~\cite{chen2013battling} study the behavior of hidden paid posters on OSNs. 
To better understand how these actors work, an author of this work posed as a hidden paid poster for a site\cite{shuijunwang} that gives users the option to be hidden paid posters.
This task revealed valuable information regarding the organization of such sites and the behavior of the hidden paid posters, who are assigned with missions that need to be accomplished within a deadline. 
For example, a mission can be about posting articles of a particular content on different sites.
A manager of the site can verify the completion of the task and then the hidden paid poster gets paid.
To further study the problem, they collect data ,pertaining to a dispute between two big Chinese IT companies, from users of 2 popular Chinese news sites (namely Sohu \cite{sohu} and Sina \cite{sina}). 
During this conflict there were strong suspicions that both companies employed hidden paid posters to disseminate false information that aimed to inflict damage to the other company.  
By undertaking statistical and semantic analysis on the hidden paid posters' content they uncover a lot of useful features that can be used in identifying hidden paid posters.
To this end, they propose the use of SVMs in order to detect such users by taking into consideration statistical and semantic features; their evaluation show that they can detect users with 88\% accuracy.

\descr{Rumors.} Starbird et al.~\cite{starbird2016could} study and identify various types of expressed uncertainty within posts in OSN during a rumor's lifetime.
To analyze the uncertainty degree in messages, the paper acquires 15M tweets related to two crisis incidents (Boston Bombings and Sydney Siege). 
They find that specific linguistic patterns are used in rumor-related tweets.
Their findings can be used in future detection systems in order to detect rumors effectively in a timely manner.
Zubiaga et al.~\cite{zubiaga2015towards} propose a different approach in collecting and preparing datasets for false information detection. 
Instead of finding rumors from busting websites and then retrieving data from OSNs, they propose the retrieval of OSN data that will subsequently annotated by humans.
In their evaluation, they retrieve tweets pertaining to the Ferguson unrest incident during 2014.
They utilize journalists that act as annotators with the aim to label the tweets and their conversations. Specifically, the journalists annotated 1.1k tweets, which can be categorized into 42 different stories. Their findings show that 24.6\% of the tweets are rumorous.
FInally, Spiro et al.~\cite{spiro2012rumoring} undertake a quantitative analysis on tweets pertaining to the 2010 Deepwater Horizon oil spill. 
They note that media coverage increased the number of tweets related to the disaster. 
Furthermore, they observe that retweets are more commonly transmitted serially when they have event-related keywords.

\subsection{Systems}
\descr{Biased.} Park et al.~\cite{park2009newscube} note that biased information is profoundly disseminated in OSNs.
To alleviate this problem, they propose NewsCube: a service that aims to provide end-users with all the different aspects of a particular story. 
In this way, end-users can read and understand the stories from multiple perspectives hence assisting in the formulation of their own unbiased view for the story.
To achieve this, they perform structure-based extraction of the different aspects that exist in news stories. 
These aspects are then clustered in order to be presented to the end-users.
To evaluate the effectiveness of their system, they undertake several user studies that aim to demonstrate the effectiveness in terms of the ability of the users to construct balanced views when using the platform. 
Their results indicate that 16 out of 33 participants stated that the platform helped them formulate a balanced view of the story, 2 out of 33 were negative, whereas the rest were neutral.

\descr{Credibility Assessment.} Hassan et al.~\cite{hassan2014data} propose FactWatcher, a system that reports facts that can be used as leads in stories. 
Their system is heavily based on a database and offers useful features to it's users such as ranking of the facts, keyword-based search and fact-to-statement translation.
Ennals et al.~\cite{ennals2010highlighting} describe the design and implementation of Dispute Finder, which is a browser extension that allows users to be warned about claims that are disputed by sources that they might trust.
Dispute Finder maintains a database with well-known disputed claims which are used to inform end-users in real-time while they are reading stories. 
Users are also able to contribute to the whole process by explicitly flagging content as disputed, or as evidence to dispute other claims.
In the case of providing evidence, the system requires a reference to a trusted source that supports the user's actions, thus ensuring the quality of user's manual annotations.
Mitra and Gilbert~\cite{mitra2015credbank} propose CREDBANK that aims to process large datasets by combining machine and human computations. 
The former is used to summarize tweets in events, while the latter is responsible for assessing the credibility of the content.
Pirolli et al.~\cite{pirolli2009so} focus on Wikipedia and develop and system that presents users an interactive dashboard, which includes the history of article content and edits.
The main finding is that users can better judge the credibility of an article, given that they are presented with the history of the article and edits through an interactive dashboard.

\subsection{Use of images on the false information ecosystem}

Information can be disseminated via images on the Web. The use of images increases the credibility of the included information, as users tend to believe more information that is substantiated with an image.
However, nowadays, images can be easily manipulated, hence used for the dissemination of false information. 
In this section, we provide an overview of the papers that studied the problem of false information on the Web, while considering images.

\descr{Fabricated.} Boididou et al.~\cite{boididou2014challenges, boididou2015verifying} focus on the use of multimedia in false information spread in OSNs. 
In~\cite{boididou2015verifying} they prepare and propose a dataset of 12K tweets, which are manually labeled as fake, true, or unknown.
A tweet is regarded as true if the image is referring to a particular event and fake if the image is not referring to a particular event. 
The authors argue that this dataset can help researchers in the task of automated identification of fake multimedia within tweets. 
In~\cite{boididou2014challenges} they study the challenges that exist in providing an automated verification system for news that contain multimedia.
To this end, they propose the use of conventional classifiers with the aim to discern fake multimedia pertaining to real events. 
Their findings demonstrate that generalizing is extremely hard as their classifiers perform poorly (58\% accuracy) when they are trained with a particular event and they are tested with another. 
Diego Saez-Trumper~\cite{saez2014fake} proposes a Web application, called Fake Tweet Buster, that aims to warn users about tweets that contain false information through images or users that habitually diffuse false information. 
The proposed approach is based on the reverse image search technique (using Google Images) in order to determine the origin of the image, its age and its context.
Furthermore, the application considers user attributes and crowdsourcing data in order to find users that consistently share tweets that contain false information on images.
Pasquini et al.~\cite{pasquini2015towards} aim to provide image verification by proposing an empirical system that seeks visually and semantically related images on the web.
Specifically, their system utilizes news articles metadata in order to search, using Google's search engine, for relevant news articles. 
These images are then compared with the original's article images in order to identify whether the images were tampered.
To evaluate their approach, they created dummy articles with tampered images in order to simulate the whole procedure.

Jin et al.~\cite{jin2016novel} emphasize the importance of images in news articles for distinguishing its truthfulness. 
They propose the use of two sets of features extracted from images in conjunction with features that are proposed by~\cite{castillo2011information,kwon2013prominent}. 
For the image features, they define a set of visual characteristics as well as overall image statistics.
Their data is based on a corpus obtained from the Sina Weibo that comprises 50K posts and 26K images. 
For evaluating the image feature set, they use conventional machine learning techniques: namely SVM, Logistic Regression, KStar, and Random Forest.
They find that the proposed image features increase the accuracy by 7\% with an overall accuracy of 83\%. 
In a follow-up work, Jin et al.~\cite{jin2016image} leverage deep neural networks with the goal of distinguishing the credibility of images. 
They note that this task is extremely difficult as images can be misleading in many ways.
Specifically, images might be outdated (i.e., old images that are falsely used to describe a new event), inaccurate, or even manipulated.
To assess the image credibility, they train a Convolutional Neural Network (CNN) using a large-scale auxiliary dataset that comprises 600K labeled fake and real images.
Their intuition is that the CNN can extract useful hyperparameters that can be used to detect eye-catching and visually striking images, which are usually used to describe false information. 
Their evaluation indicates that the proposed model can outperform several baselines in terms of the precision, recall, F1, and accuracy scores.
Gupta et al.~\cite{gupta2013faking} focus on the diffusion of fake images in Twitter during Hurricane Sandy in 2012. 
They demonstrate that the use of automated techniques (i.e., Decision Trees) can assist in distinguishing fake images from real ones. 
Interestingly, they note that the 90\% of the fake images came from the top 0.3\% of the users.

\section{Discussion \& Conclusions}
\label{sec:conclusions}

In this work, we have presented an overview of the false information ecosystem. 
Specifically, we have presented the various types of false information that can be found online, the different actors of the false information ecosystem as well as their motives for diffusing controversial information.
Through the identification of several lines of work, we have presented the existing work on the false information ecosystem. 
Namely, we have presented studies on user perception, propagation dynamics, detection and containment of false information, as well as the dynamics of false information on the political stage.
Also, we present some gaps of the existing literature that can be exploited by researchers in order to further study the increasing problem of false information on the Web.

To conclude, we share some thoughts about the  problem of false information on the Web's ecosystem. 
We argue that at the current stage current automated solutions, that do not use human input, are unable to effectively mitigate the problem of false information on the Web. 
Therefore, we feel that we should put extra effort in raising awareness of the problem to regular users of social networks, so that they can distinguish false information and potentially understand if a post is made from a legitimate user instead of a bot or state-sponsored actors.
When it comes to scientific research and solutions for the problems, we argue that it is extremely important to tackle the problem from a holistic point of view. 
That is, researchers should take into account multiple Web communities when considering the problem of false information on the Web.
Also, we should focus on providing models and methods that generalize well on other communities or datasets.
Finally, researchers should focus on designing and developing real-time platforms that will shed light about the propagation of false information across multiple Web communities.
For instance, inform Twitter users that 4chan users are pushing a particular ``hashtag'' in the Twitter platform, with the goal of promoting information of questionable credibility.

\section{Acknowledgments}
This work is supported by the European Union's Horizon 2020 research and innovation programme under the Marie Sklodowska-Curie ``ENCASE'' project (Grant Agreement No. 691025).

\small
\bibliography{bibliography} 
\bibliographystyle{abbrv}

\end{document}